\documentclass[letter]{aa} 

\usepackage{graphicx}
\usepackage{txfonts}
\usepackage[colorlinks,citecolor=blue]{hyperref}
\hypersetup{
    colorlinks = true,
    linkcolor = blue,
    anchorcolor = blue,
    citecolor = blue,
    filecolor = blue,
    urlcolor = blue
    }

\usepackage{mathrsfs}
\usepackage{multirow,bm}
\usepackage{svg}
\usepackage{ulem}

\setcitestyle{notesep={},round,aysep={},yysep={;}}

\begin{document} 

   \title{Runaway BN supergiant star HD 93840: Progenitor of an imminent core-collapse supernova above the Galactic plane}

   \author{D. We{\ss}mayer\inst{1}
          \and
          M. A. Urbaneja\inst{1}
          \and
          K. Butler\inst{2}
          \and
          N. Przybilla\inst{1}
          }

   \institute{Universit\"at Innsbruck, Institut f\"ur Astro- und Teilchenphysik, Technikerstr. 25/8, 6020 Innsbruck, Austria\\
              \email{david.wessmayer@uibk.ac.at ; norbert.przybilla@uibk.ac.at}
         \and
             Ludwig-Maximilians-Universit\"at M\"unchen, Universit\"atssternwarte, Scheinerstr. 1, 81679 M\"unchen, Germany
             }

   \date{Received ; accepted }

 
  \abstract{We present a quantitative spectral analysis of the extreme nitrogen-enhanced supergiant HD 93840 (BN1\,Ib) at an intermediate galactic latitude. Based on an optical high-resolution spectrum and complementary ultraviolet and infrared (spectro-)photometry, in addition to \textit{Gaia} data, we carried out a full characterisation of the star's properties. We used both hydrostatic and unified (photosphere+wind) model atmospheres that account for deviations from local thermodynamic equilibrium. A highly unusual surface CNO-mixing signature and a marked stellar overluminosity compared to the mass imply a binary channel for the star's past evolution. The kinematics shows that it has reached its current position above the Galactic plane as a runaway star, likely ejected by the supernova explosion of its former companion star. Its current bulk composition, with a notably increased mean molecular weight due to core He- and progressed shell H-burning, suggests an advanced evolutionary stage. It is poised to  yield a rare core-collapse supernova of a blue supergiant about ten OB star population scale heights above the Galactic disk relatively soon, contributing to the metal enrichment of the circumgalactic~medium.}

   \keywords{Stars: abundances -- Stars: atmospheres -- Stars: evolution -- Stars: fundamental parameters -- supergiants
               }

   \authorrunning{We{\ss}mayer et al.}
   \titlerunning{ The runaway BN supergiant star HD 93840}

   \maketitle

%

\section{Introduction} \label{sec:intro}
Core-collapse supernovae (ccSN) from massive stars ($M$\,$\gtrsim$\,8\,$M_\odot$) play an important role in the cosmic matter cycle by generating metals and injecting them into the interstellar medium (ISM) rapidly on cosmic timescales. As massive stars are formed in clusters and OB associations, the collective effects of multiple spatially and temporally ($\sim$10$^7$\,yr) close ccSN lead to the formation of superbubbles in the ISM and to a breakout of hot gas into the galactic halo via galactic chimneys \citep[e.g.][]{Normandeauetal96}. This 'galactic fountain' model also leads to a distribution of metals into the circumgalactic environment. After some retention time ($\sim$10$^8$\,yr), these metals fall back onto the galactic disk and mix with the ISM over large distances \citep[][]{Tenorio-Tagle96}.

Individual ccSNe may contribute to this via in situ injections of metals high above the galactic plane. These may stem from the massive members among the runaway stars \citep[e.g.][]{SiNa11} or from  massive hyper-runaway \citep[e.g.][]{Przybillaetal08c} or hypervelocity stars \citep[e.g.][]{Przybillaetal08d}. In particular, the latter two may potentially explain the presence of SN remnants at large distances from the galactic plane \citep[][]{Filipovicetal22}, apart from SNe\,Ia. The unhindered expansion of such SNe may contribute to the metallicity buildup of the intergalactic medium.
The  massive stars observed away from the Galactic disk are typically early B-type main-sequence and giant stars because of lifetime considerations. The binary fraction of runaway stars is significantly suppressed compared to the normal OB-star population \citep{GiBo86}. As a consequence, from a naive point of view, the majority of these stars should result in SNe II-P from red supergiant progenitors \citep{Smartt09}. 

Up to now, the blue supergiant \object{HD 93840} at intermediate galactic latitude ($b$\,=\,+11.1\degr) has mostly attracted interest as a distant but nonetheless relatively bright ($V$\,=\,7.76) background source for studies of the ISM above the Galactic plane. A wide range of studies in the optical and with  International Ultraviolet Explorer \citep[IUE, e.g.][]{SaMa87},  Orbiting and Retrievable Far and Extreme Ultraviolet Spectrometer \citep[ORFEUS, e.g.][]{Widmannetal98}, and  Far Ultraviolet Spectroscopic Explorer \citep[e.g.][]{Dixonetal06} were conducted in this context. However, the supergiant itself has so far been of limited interest. After an initial report of it being overabundant in nitrogen from its UV spectrum \citep{SaMa87}, it was confirmed as an extreme N-enhanced supergiant and optically classified as BN1\,Ib by \citet{Walbornetal90}. \citet{Massaetal91} derived surface carbon to nitrogen abundances of 1:10 from a differential (model-atmosphere independent) analysis of the UV wind lines of HD~93840, whereas the oxygen abundance appeared to be normal. They concluded that the photospheric material was only once shortly exposed to CN-burning, avoiding the slower modification by the ON-cycle. A single analysis with a modern model atmosphere code provided limited information \citep{Fraser_etal_10}, as seen in Table~\ref{tab:parameters}. Here, we present a comprehensive analysis that shows that HD~93840 is far more interesting beyond what is known so far, constituting a unique highly-evolved and markedly overluminous runaway star that will likely explode in a rare ccSN of a blue supergiant within a relatively short timescale and located well above the Galactic plane. 

\section{Observational data}
Our analysis is based on two spectra of HD~93840, observed consecutively in the night of 24 April 2005 with the 
Fibrefed Extended~Range Opti\-cal Spectrograph \citep[{FEROS},][]{Kauferetal99} on the Max-Planck-Gesellschaft/European 
Southern Observatory (ESO) 2.2\,m telescope at La Silla in Chile. The phase~3 data, which were downloaded from the ESO Science 
Portal\footnote{\rule{-0.25mm}{0mm}\url{https://archive.eso.org/scienceportal/home}}, cover a useful wavelength range from about
3700 to 9200\,{\AA} at a resolving power $R$\,=\,$\lambda / \Delta \lambda$\,$\approx$\,48\,000. The spectra were normalised by fitting a spline function through carefully selected continuum points and co-added after verification of the absence of any significant radial velocity shift, resulting in a signal-to-noise ratio of $S/N$\,=\,425 per pixel, measured at 5585\,{\AA}. 

Various sources of (spectro-)photometric data were considered in this work. In the UV, we employed spectra taken with the T\"ubingen Echelle Spectrograph (TUES\footnote{\url{https://archive.stsci.edu/tues/}}) on board ORFEUS-SPAS II \citep{Barnstedt_etal_99}. Additionally, we used high-dispersion, large-aperture spectra (data IDs SWP49711 and LWP27124) taken by IUE\footnote{\url{https://archive.stsci.edu/iue/}}. 
We also adopted UV-photometry from the Belgian/UK Ultraviolet Sky Survey 
Telescope \citep[S2/68,][]{Thompson95} on board the European Space Research Organisation TD1 satellite. In the optical, we considered Johnson $UBV$ magnitudes \citep{Mermilliod97} as well as low-resolution spectra of \textit{Gaia} Data Release 3 \citep[DR3,][]{Gaia2016,GaiaDR3}. Finally, in the IR wavelength range, we employed 2MASS $JHK$ magnitudes \citep{Cutrietal03} and Wide-Field Infrared Survey Explorer (WISE) photometry \citep{Cutrietal21}.

\section{Analysis methodology}

Modelling the photospheric spectra of early B-type supergiants requires the consideration of deviations from local thermodynamic equilibrium (so-called non-LTE effects) and (in cases of very high luminosity) an explicit treatment of the stellar wind and spherical extension. While HD~93840 at luminosity class Ib is clearly not highly luminous, we nevertheless adopted two independent modelling approaches for verification purposes.

In a hybrid non-LTE approach, model atmospheres were calculated with \,{\sc Atlas12} \citep{kurucz05} as line-blanketed, plane-parallel, homogeneous, and hydrostatic structures in LTE. This provided the basis for non-LTE line-formation calculations carried out with updated and extended versions of {\sc Detail} and {\sc Surface} \citep{Giddings81,BuGi85} and state-of-the-art model atoms. Grids of synthetic spectra were calculated and matched to the observations using $\chi^2$-minimisation, employing the Spectral Plotting and Analysis Suite \citep[{\sc Spas},][]{Hirsch09}. Multiple spectroscopic indicators were utilised to provide tight constraints on the atmospheric parameters. For details of the analysis procedure, we refer to \citet[, henceforth Papers~I \& II]{Wessmayeretal22,Wessmayeretal23} and we also include a brief summary in Appendix~\ref{appendix:A}. The analysis technique based on the modelling via \,{\sc Atlas12}\,+ {\sc Detail}\,+\,{\sc Surface} (abbreviated as {\sc Ads} henceforth) was shown to be applicable to a wide range of massive stars, that is, early B- \citep{NiPr07,NiPr12} and late O-type main-sequence stars \citep{Aschenbrenneretal23}, as well as BA-type supergiants \citep[, Papers I, II]{Przybillaetal06,FiPr12}.

As a second option, unified (wind+photosphere) non-LTE model atmospheres were computed with {\sc Fastwind} \citep{Pulsetal05,Pulsetal20}, which accounts for non-LTE line blanketing, mass outflow, and spherical geometry. The code has been extensively used for the analysis of B-type supergiants in the Milky Way \citep[e.g.][]{Herreroetal22} as well as in other galaxies \citep[e.g.][]{Urbanejaetal05a,Urbanejaetal17,Kudritzkietal16}. 

The interstellar sightline towards HD~93840 was characterised using the mean extinction law of \citet{fitzpatrick99}. For this, the spectral energy distribution (SED) of the {\sc Atlas} model was reddened and fitted to the observed (spectro-)photometry for the two parameters of colour excess $E(B-V)$ and total-to-selective extinction $R_V$\,=\,$A_V$/$E(B-V)$. The interstellar extinction $A_V$ was then determined as the product of the two values.

\begin{table}
\caption{Atmospheric and stellar parameters of HD~93840.}
\vspace{-0.6cm}
\label{tab:parameters}
{\footnotesize
\setlength{\tabcolsep}{1.1mm}
\begin{center}    
\begin{tabular}{cccc}        
\hline\hline
\multicolumn{4}{l}{General information:} \\
Sp. Type & BN1\,Ib &  $\varpi$\,(mas)\tablefootmark{a} & $0.3604 \pm 0.0348$\\
$\varv_{\mathrm{rad}}$\,(km\,s$^{-1}$) & $-8.1 \pm 0.5$ & $\mu_{\alpha}$\,(mas\,yr$^{-1}$)\tablefootmark{b} & $-5.767 \pm 0.03$\\
$d$\,(kpc)\tablefootmark{c} & $\sim$2.6 to $\sim$2.8 & $\mu_{\delta}$\,(mas\,yr$^{-1}$)\tablefootmark{b} & $5.033 \pm 0.03$\\
$V$\,(mag)\tablefootmark{d} & $7.761 \pm 0.013$ & $B-V$\,(mag)\tablefootmark{d} & $-0.048 \pm 0.008$\\ [3.5mm]
\multicolumn{4}{l}{Stellar parameters and elemental abundances:} \\
                           & {\sc Ads}\tablefootmark{e} & {\sc Fastwind}\tablefootmark{f} & {\sc Tlusty}\tablefootmark{g} \\ \hline
$T_{\mathrm{eff}}$\,(kK)         &  $21.8 \pm 0.3$    &  $23.1^{+0.9}_{-1.0}$     & $20.9 \pm 1.0$\\
$\log g$\,(cgs)            & $3.00 \pm 0.05$     &  $2.90^{+0.08}_{-0.1}$ & $2.75 \pm 0.2$\\
$\xi$\,(km\,s$^{-1}$)            & $13 \pm 2$     &  $16 \pm 3$     & $15 \pm 3$ \\
$\varv  \sin i$\,(km\,s$^{-1}$)  &  $68 \pm 3$    &  $68$ & $58 \pm 3$\\
$\zeta$\,(km\,s$^{-1}$)            &  $65 \pm 5$    & $65$ & $68 \pm 4$\\ \hline
$\varepsilon$(He)           &  $11.06^{+0.07}_{-0.09}$    &  $11.26^{+0.18}_{-0.22}$ &  ...\\
$\varepsilon$(C)            &  $7.25 \pm 0.09$    &  $7.29^{+0.18}_{-0.16}$  & ...\\
$\varepsilon$(N)            &  $8.59 \pm 0.08$    &  $8.53^{+0.10}_{-0.09}$  & $8.37 \pm 0.2$\\
$\varepsilon$(O)            &  $8.42 \pm 0.05$    &  $8.37^{+0.11}_{-0.10}$  & ...\\
$\varepsilon$(Ne)           & $8.10 \pm 0.07$    &  ...     & ...\\
$\varepsilon$(Mg)           &  $7.40 \pm 0.05$    &  $7.44^{+0.13}_{-0.12}$  & ...\\
$\varepsilon$(Al)           & $6.24 \pm 0.08$     &  ...     & ...\\
$\varepsilon$(Si)           &  $7.63 \pm 0.08$    &  $7.33^{+0.12}_{-0.11}$  & ...\\
$\varepsilon$(S)            &  $7.32 \pm 0.05$    &  ...     & ...\\
$\varepsilon$(Ar)           & $6.61 \pm 0.08$     &  ...     & ...\\
$\varepsilon$(Fe)           &  $7.33 \pm 0.08$    &  ...     & ...\\ 
$Z$            & $0.011 \pm 0.002$     & ...      & ...\\ \hline
 $R_V$ & $2.9 \pm 0.1$ & $ 2.95 \pm 0.35$ & ...\\
 $~~E\left(B-V\right)$\,(mag) & $0.21 \pm 0.03$ &  $0.19 \pm 0.02$ & ...\\
$M_V$\,(mag) & \multicolumn{2}{c}{$-$4.82 to $-$5.20} & ...\\
$M_{\mathrm{bol}}$\,(mag) & \multicolumn{2}{c}{$-$6.85 to $-$7.23} & ...\\
$M/M_{\odot}$ & \multicolumn{2}{c}{7.8 to 11.0} & $23\pm5$\\
$R/R_{\odot}$ & \multicolumn{2}{c}{14 to 17} & ...\\
$\log L/L_\sun$ & \multicolumn{2}{c}{4.63 to 4.79} & ...\\
$\log \tau/\mathrm{yr}$ & \multicolumn{2}{c}{$>$7.10} & ...\\[0.3mm]
\hline
\end{tabular}
\end{center}}
\vspace{-0.2cm}
\tablefoot{Uncertainties are 1$\sigma$-standard deviations, except where noted otherwise. 
\tablefoottext{a}{\cite{GaiaDR3}} 
\tablefoottext{b}{Accounting for magnitude-dependent systematics of \textit{Gaia} EDR3 data according to \citet{CGB21}.}
\tablefoottext{c}{Most likely full range, see Appendix~\ref{appendix:B}.}
\tablefoottext{d}{\cite{Mermilliod97}}
\tablefoottext{e}{Based on the {\sc Atlas12+Detail+Surface (Ads)} solution.} 
\tablefoottext{f}{Derived using {\sc Fastwind}.}
\tablefoottext{g}{\cite{Fraser_etal_10}.} 
For the fundamental parameters full ranges based on the likely distance interval are given.
}
\end{table}

\begin{figure}[ht!]
\centering
\includegraphics[width=.99\linewidth]{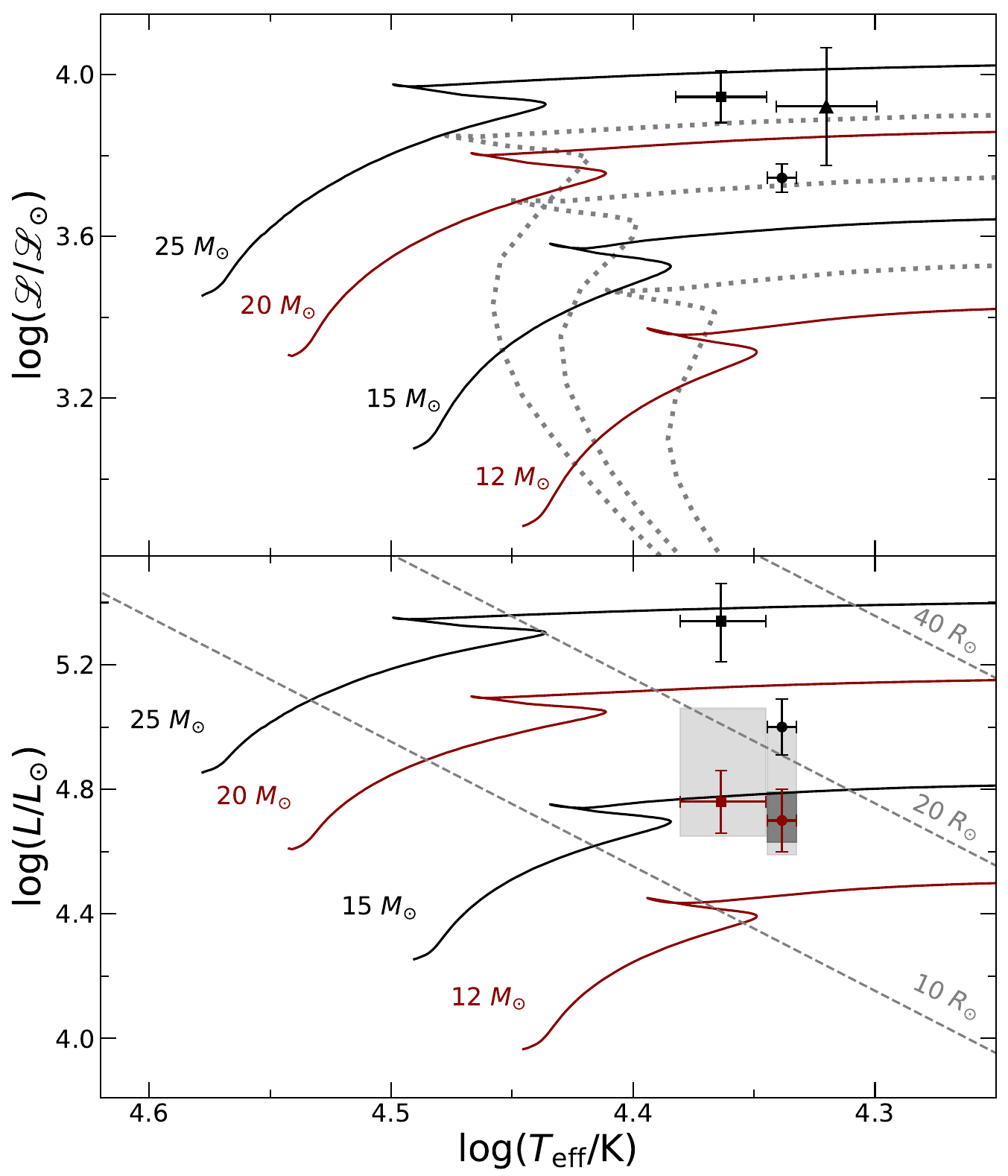}
\caption{Location of HD~93840 in two diagnostic diagrams, the sHRD (upper panel) and the HRD (lower panel). Parameters derived via {\sc Ads} (dots), {\sc Fastwind} (squares), and {\sc Tlusty} (triangle) considering the spectroscopic and \textit{Gaia}-based distances (i.e. $d_\mathrm{spec}$ and $d_\mathrm{Gaia}^\mathrm{MA,EB}$, see Appendix~\ref{appendix:B}), are depicted as black and red symbols, respectively. The lightgrey boxes span the parameter range obtained when accounting in addition to parallax bias according to \citet{MaizApellaniz22}, while the darkgrey box represents the finally adopted solution from Table~\ref{tab:parameters}. For comparison, loci of evolution tracks for stars rotating at 
$\Omega_{\mathrm{rot}}$\,=\,0.568\,$\Omega_{\mathrm{crit}}$ \citep{Ekstroemetal12} are indicated for various zero-age
main-sequence (ZAMS) masses. Isochrones for the model grid, corresponding to ages of $\log \tau_\mathrm{evol} \in \{6.95, 7.05, 7.20\}$ are shown as dotted lines in the upper panel (with age increasing from top to bottom). Error bars indicate 1$\sigma$ uncertainty ranges. See the text for a discussion.
\label{fig:hrd}}
\end{figure}

Once the atmospheric parameters and the interstellar extinction are determined, the fundamental stellar parameters depend only on the distance. Various values may be employed in view of the different bias corrections proposed for the \textit{Gaia} measurements, the spectroscopic distance may be adopted. Further constraints on the distance, based on the runaway nature of HD~93840 and its advanced evolutionary state, may be made (this is summarised in Appendix~\ref{appendix:B}). A fixed distance constrains the absolute visual magnitude, $M_V$, and with aid of the {\sc Atlas} model bolometric correction the absolute bolometric magnitude, $M_\mathrm{bol}$. This can be converted into luminosity, $L$, which provides the stellar radius, $R$, for a given $T_\mathrm{eff}$. Finally, from $R$ and $\log g$, we  can calculate the stellar mass, $M$. Usually, the stellar age, $\tau$, can be determined from isochrones, which is inappropriate in this case, as the star's evolution is not described well by the associated evolution tracks. Instead, a lower age limit is derived from the flight time from the Galactic mid-plane to the current location of HD~93840 above the disk (see Appendix~\ref{appendix:C} for details).

\begin{figure}[ht!]
\centering
\includegraphics[width=.995\linewidth]{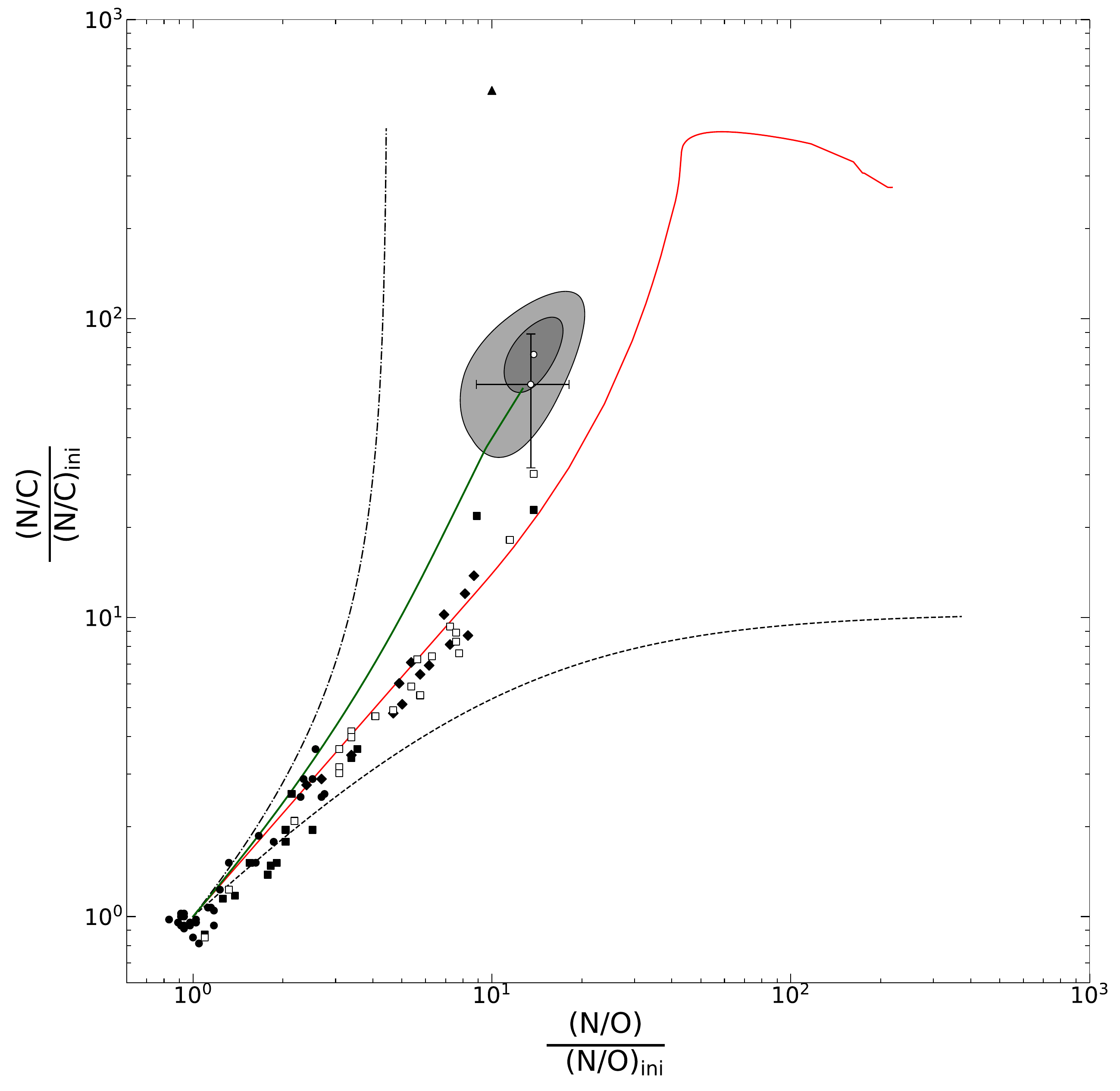}
\caption{Nitrogen-to-carbon ratio vs.~nitrogen-to-oxygen ratio, normalised to initial values. The mixing ratios derived for HD~93840 are depicted as open dots with grey 1$\sigma$ and 2$\sigma$ uncertainty regions ({\sc Ads}) and with error bars depicting the 1$\sigma$ uncertainty ({\sc Fastwind}). Objects from our previous work are indicated: B-type main-sequence stars \citep[][, black dots]{NiSi11,NiPr12}, late O-type main-sequence stars \citep[, black squares]{Aschenbrenneretal23}, BA-type supergiants \citep[][, black diamonds]{Przybillaetal10}, B~supergiants (Paper I \& II, open squares), and the stripped CN-cycled core $\gamma$ Columbae \citep[, black triangle]{Irrgangetal22}.
For comparison, the development of the surface CNO abundances is shown for
a 25\,$M_\sun$, $\Omega_{\mathrm{rot}}$\,=\,0.568\,$\Omega_{\mathrm{crit}}$ model \citep[][{, red}]{Ekstroemetal12} and for a 15\,$M_\sun$, $\Omega_{\mathrm{rot}}$\,=\,0.95\,$\Omega_{\mathrm{crit}}$ model \citep[][, green line]{Georgyetal13}. The dashed and dash-dotted lines depict the analytical boundaries for the ON- and CN-cycle, respectively. A discussion is provided below.
\label{fig:cno}}
\end{figure}

\section{Results} \label{sec:results}
All results from the quantitative analysis based on {\sc Ads} and {\sc Fastwind}, plus the data obtained from the use of the {\sc Tlusty} code by \citet{Fraser_etal_10} are summarised in Table~\ref{tab:parameters}. Overall, there is agreement seen among the derived atmospheric parameters of the three solutions within the mutual error bars, with the {\sc Tlusty} analysis tending to find a lower $T_\mathrm{eff}$, whereas the {\sc Fastwind} solution indicates a higher $T_\mathrm{eff}$ (the latter follows a general trend, see Papers I \& II). Line broadening parameters due to rotation and macroturbulence also agree well, except for a slightly smaller $\varv \sin i$-value in the {\sc Tlusty} solution. The trend in the microturbulent velocity found in the {\sc Ads} solution is towards smaller values than in the other solutions, in accordance with the findings of Papers I \& II. This has consequences for the derivation of metal abundances which should be lower at higher $\xi$-values, in particular for chemical species where the abundance determination mostly relies on stronger lines (e.g. for silicon). This is exactly what we found here. The restricted set of elemental abundances derived using {\sc Fastwind} is due to the lack of detailed model atoms, and elemental abundances besides those of nitrogen were not investigated by \citet{Fraser_etal_10}. We  mostly concentrate on the {\sc ads} solution for further interpretation, because of the smaller uncertainties that arise from the use of more complex model atoms and the consideration of a larger number of observational constraints in the spectroscopic analysis than employed in the other approaches.

The comparison of our final {\sc Ads} and {\sc Fastwind} synthetic spectra with the full observations is shown in Figs.~\ref{fig:hd93840_1} to ~\ref{fig:hd93840_9}. Overall, an excellent fit was achieved, implying that the detailed photospheric stratification is well-matched by the models (see Appendix~\ref{appendix:A} for a further discussion). A high level of quality was also achieved for the SED fit from $\sim$0.1 to 20\,$\mu$m as shown in Fig.~\ref{fig:sed}. Therefore, the global energy output of the star was  also reliably reproduced (see also Appendix~\ref{appendix:A} for further details).

\begin{figure*}[ht!]
\sidecaption
\includegraphics[width=12cm]{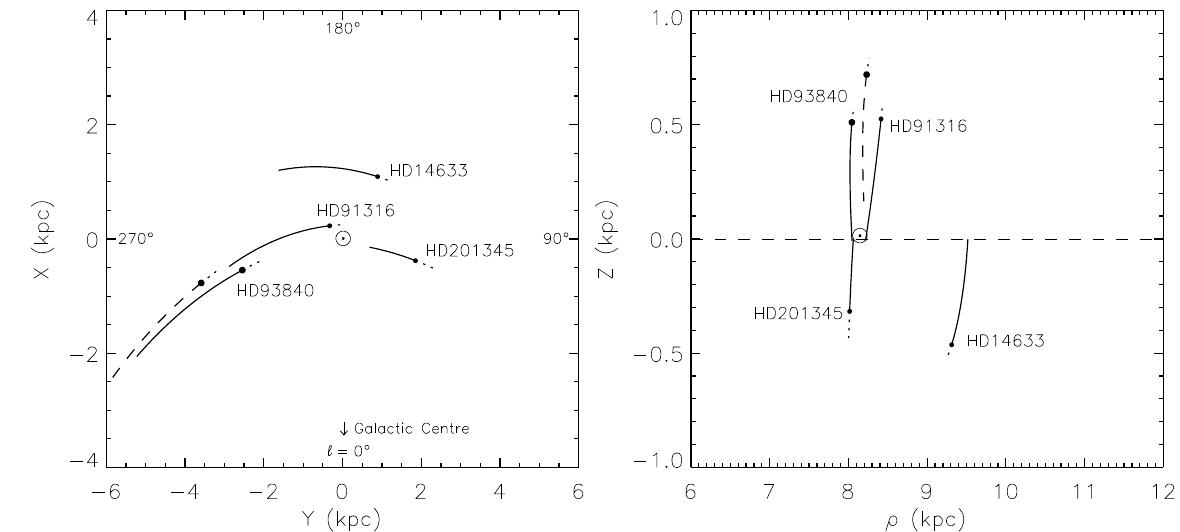}
\caption{Kinematics of HD~93840 (full line: adopting the \textit{Gaia}-based distance $d_\mathrm{Gaia}^\mathrm{MA,EB}$, see Appendix~\ref{appendix:B}; dashed: spectroscopic distance $d_\mathrm{spec}$) in the Galactic potential, together with several other runaway stars from our previous work. Galactic Cartesian coordinates $XYZ$ are employed, with the origin shifted to the position of the Sun; $\rho$ is the galactocentric distance. Displayed here: the Galactic plane projection (left panel) and the movement in the meridional plane (right panel). The dots mark the current positions and the orbits are traced back for the stellar age (full lines) and 2\,Myr into the future (dotted lines). See the text for a discussion.
\label{fig:kinematics}}
\end{figure*}

The position of HD~93840 in the spectroscopic Hertzsprung-Russell diagram (sHRD, $\log \mathscr{L}/\mathscr{L}_\odot$ vs. $\log T_\mathrm{eff}$, $\mathscr{L}$\,=\,$T_\mathrm{eff}^4/g$) and the classical HRD is shown in Fig.~\ref{fig:hrd}. The differences in the atmospheric parameters, which agree well overall, become amplified by the combination into $\mathscr{L}$, such that the three solutions from Table~\ref{tab:parameters} span quite a range in the parameter space of the sHRD. When comparing the positions of the {\sc Ads} and the {\sc Fastwind} solutions in the sHRD and those based on the spectroscopic distance in the HRD, a good agreement is found. However, the \textit{Gaia}-based luminosities (the two red symbols and grey boxes, see the discussion in Appendix~\ref{appendix:B}) are notably lower, but they mutually agree much better -- except for the aforementioned shift in $T_\mathrm{eff}$. 

The determination of the mass and stellar age by comparison with evolution tracks and isochrones in the sHRD apparently has the advantage that only knowledge of $T_\mathrm{eff}$ and $\log g$ is required. However, there is a catch: the evolutionary tracks, such as the ones for rotating stars by \citet{Ekstroemetal12} used here, have to come close to the real evolutionary scenario for the investigated star. That is, the star is a normally evolving single star or an object in a binary that has not yet experienced interaction. Usually, good agreement between the spectroscopic and \textit{Gaia} parallax-based distances has been achieved in our previous studies of massive stars \citep[see e.g. Papers I \& II,][]{Aschenbrenneretal23}. In that case, the positions of a star relative to the evolutionary tracks in the sHRD and the HRD are similar. However, there are exceptions, as is the case for HD~93840 in the present work \citep[and the two ON stars discussed by][]{Aschenbrenneretal23}. This is because our most likely distance value of $\sim$2.6 to 2.8\,kpc is significantly shorter than $d_\mathrm{spec}$\,$\approx$\,3.7\,kpc and, therefore, the luminosity, radius and mass of HD~93840 are lower. 
This implies that the tracks used are inappropriate to describe the evolutionary status of the star correctly. In the following, further constraints will be employed to investigate what kind of object HD~93840~really~is.

\section{Discussion}\label{sec:discussion}
The usual suspects for blue stars at high galactic latitudes with super\-giant characteristics are post-asymptotic giant branch stars, but such a scenario can be excluded for HD~93840 because pre-white dwarfs have masses smaller than the Chandrasekhar limit. We therefore resort to investigating the surface CNO abundances as indicators for the evolutionary status. The nitrogen-to-carbon ratio versus the nitrogen-to-oxygen ratio \citep[normalised to initial abundances adopted from][ and the model abundance ratios normalised to their respective initial values]{NiPr12} of HD~93840 is depicted in Fig.~\ref{fig:cno}, where it is also set into the context of massive stars we previously analysed. The mixing of the surface layers with CNO-processed matter follows a tight path in the diagram \citep{Przybillaetal10,Maederetal14}. One clear outlier was recently found, namely: the stripped star $\gamma$ Col, which exposes previously CN-burning layers on its surface \citep{Irrgangetal22}. The mixing signature of HD~93840 also deviates from the locus of the other stars. Yet, in principle, it could  be reached by a near-critically rotating single 15\,$M_\odot$ star at the end of its lifetime \citep[adopting a model of][ with an initial $v_\mathrm{rot}$\,=\,525\,km\,s$^{-1}$, the green line in Fig.~\ref{fig:cno}]{Georgyetal13}, as lower-mass stars tend to have less O-depletion, due to a slower ON-cycle relative to more massive stars. However, HD~93840 would be expected to be a red supergiant from this model\footnote{
{We note that single stars of such mass do not experience mass loss strong enough to shed sufficient hydrogen-rich envelope to turn into a blue supergiant.}}, contrary to observation -- rendering this scenario inapplicable.

The observed high mixing signature points in the right direction, however. We can compare the luminosity of HD~93840 with a normal star of the same $M$ and $T_\mathrm{eff}$ \citep[interpolating the models of][]{Ekstroemetal12} and find a notable overluminosity. The generalised mass-luminosity relationship is $L$\,$\propto$\,$M^\alpha \mu^\beta$, with $\mu$\,=\,$(2X+\frac{3}{4}Y+\frac{1}{2}Z)^{-1}$ being the mean molecular weight for mass fractions of hydrogen $X$, helium $Y$ and the metals $Z$, and $\beta$\,$\approx$\,4 \citep{Kippenhahnetal13}. The overluminosity therefore stems from a significantly increased mean molecular weight for the bulk stellar material, from about 0.6 for a standard initial composition to about 0.88 to 1.07 at present for our most likely distance range. As the mean molecular weight of H-burning regions increases to a maximum value of 1.34 when fully converted to helium, this implies the presence of a substantial He-burning core of several solar masses (including He-burning products) and an increased He-abundance in the envelope (see Appendix~\ref{appendix:B} for further discussion). With such a composition the star will stay a blue supergiant throughout its remaining life that will end in a ccSN, and a binary channel for its formation is indicated \citep[e.g.][]{Wellsteinetal01}.

Sometimes further information on the formation scenario can be derived from a star's kinematics. This is the case here and in Appendix~\ref{appendix:C}, we offer more  details on the modelling and some further discussion. The visualisation of the orbit in the disk and in the meridional plane in Fig.~\ref{fig:kinematics} shows that HD~93840 is a runaway star originating from the Carina-Sagittarius spiral arm in the fourth Galactic quadrant \citep[see e.g.][]{PantaleoniGonzalezetal21}. It exhibits a peculiar velocity perpendicular to the disk, such that it could have reached its current position within about 12.5 to 12.7\,Myr. Its current velocity in the Galactic $Z$-direction is $\sim$28\,km\,s$^{-1}$, slowed down from a value of $\sim$48\,km\,s$^{-1}$ at Galactic midplane crossing (the assumed point of binary breakup). We note that the second solution of HD~93840, based on the spectroscopic distance and assuming it to have evolved as a single star, would allow it to reach its current position within the star's lifetime when started from about 150\,pc above the disk. However, we have discarded the scenario because of the other peculiarities of the star. The footpoint of HD~93840's trajectory in the disk is close to the galactocentric distance of the Sun \citep[8.178\,kpc,][]{GravityCol1aboration19}. Thus, the metal abundances of the star should be close to cosmic abundance standard values \citep[][]{Przybillaetal08b,NiPr12}, which are representative for the present-day chemical composition of the solar vicinity. With the clear exception of CNO, this is indeed the case on the whole.

An SN ejection \citep{Blaauw61} is compatible with a binary evolution scenario, but the solid angle of HD~93840 (as seen from the SN) appeared small enough to avoid any notable pollution with SN ejecta, which is rare \citep{Przybillaetal08c,Irrgangetal10}. In fact, this, the CNO signature, and the bulk composition of HD~93840 may be used to put further constraints on the binary evolution scenario. Usually, it is the mass accretor that enters a long-lived phase in the OB-star domain, whereas the mass-losing primary star turns into a (very) hot helium star \citep[e.g.][]{Wellsteinetal01}. As the accreting secondary shows only a small overluminosity, if at all \citep{Senetal22}, such a scenario can be excluded for HD~93840. 

We sketch out the following scenario: HD~93840 was the primary of a close binary system which transferred mass in a Case A scenario to its initially lower mass-companion, thus becoming a relatively long-lived He-core, that is: a (massive) sdO subdwarf star. The secondary was spun-up by the mass transfer to close to critical velocity, became rejuvenated and its evolution became faster than that of the primary; this is a situation that is encountered in early Be-star-and-sdO subdwarf systems \citep[e.g.][]{Wangetal21}. In particular, if the mass transfer became non-conservative at some point, the separation of the two stars could have grown larger. Mass transfer from the evolved secondary in an early Case B scenario (i.e. after termination of core H-burning) could then have transferred back strongly CNO-processed material to the He-core primary. The SN explosion of the secondary's core shortly after then disrupted the system, sending HD~93840 on its observed trajectory. As the chemically most-processed material was deposited on the surface, this created an inverse $\mu$-gradient in the radiative envelope of HD~93840. Thermohaline mixing set in and led to a dilution of the initial CNO signature, similar to what would otherwise be reached by convectively mixing highly CNO-processed material from near the core with the envelope of a red supergiant. This scenario should certainly be substantiated by dedicated binary evolution calculations, however, this is beyond the scope of the present work. 

Irrespective of the details of the binary scenario, with an evolutionary stage advanced well into He-burning, a He-core mass above the Chandrasekhar limit, and the high overall mean molecular weight keeping the star from evolving to red supergiant dimensions, the core-collapse of HD~93840 as a blue supergiant seems imminent within a relatively short timescale. The resulting SN explosion may share some similarities with SN1987A due to its blue supergiant progenitor \citep{Walbornetal89} as well as  differences due to environment and progenitor evolution \citep[e.g.][]{Podsiadlowski92}. We note that our estimated He-core mass of about 4 to 5\,$M_\odot$ (see Appendix~\ref{appendix:B}) is a bit lower than that of the SN1987A progenitor \citep[5--7\,$M_\odot$,][]{Arnettetal89}, while the envelope mass of HD~93840 is significantly lower. The ccSN of HD~93840 will occur about 500 to 600\,pc (i.e. about ten typical OB star population scale heights) above the Galactic mid-plane. By injecting metals into the circumgalactic medium, it will feed the metallicity buildup of the circumgalactic medium more directly than via galactic fountains.

\begin{acknowledgements}
We express our gratitude to U.~Bastian for invaluable advice regarding \textit{Gaia} data. We also want to thank the referee for suggestions that helped to improve the paper and our second reviewer for clarifications and constructive
propositions on considering parallax bias for the distance determination.
D.W. and N.P. gratefully acknowledge support from the Austrian Science Fund FWF project DK-ALM, grant W1259-N27. Based on observations collected at the European Southern Observatory under ESO programmes 075.D-0103(A) and 092.C-0173(A), obtained from the European Southern Observatory Science Archive Facility with DOIs \url{https://doi.org/10.18727/archive/24} and \url{https://doi.org/10.18727/archive/50}.
\end{acknowledgements}


%
%

\typeout{}
\bibliographystyle{aa}
\bibliography{biblio.bib}

\begin{thebibliography}{77}
\expandafter\ifx\csname natexlab\endcsname\relax\def\natexlab#1{#1}\fi

\bibitem[{{Allen} \& {Santillan}(1991)}]{AlSa91}
{Allen}, C. \& {Santillan}, A. 1991, \rmxaa, 22, 255

\bibitem[{{Arnett} {et~al.}(1989){Arnett}, {Bahcall}, {Kirshner}, \&
  {Woosley}}]{Arnettetal89}
{Arnett}, W.~D., {Bahcall}, J.~N., {Kirshner}, R.~P., \& {Woosley}, S.~E. 1989,
  \araa, 27, 629

\bibitem[{{Aschenbrenner} {et~al.}(2023){Aschenbrenner}, {Przybilla}, \&
  {Butler}}]{Aschenbrenneretal23}
{Aschenbrenner}, P., {Przybilla}, N., \& {Butler}, K. 2023, \aap, 671, A36

\bibitem[{{Bailer-Jones} {et~al.}(2021){Bailer-Jones}, {Rybizki}, {Fouesneau},
  {Demleitner}, \& {Andrae}}]{Bailer-Jones_etal_2021}
{Bailer-Jones}, C.~A.~L., {Rybizki}, J., {Fouesneau}, M., {Demleitner}, M., \&
  {Andrae}, R. 2021, \aj, 161, 147

\bibitem[{{Barnstedt} {et~al.}(1999){Barnstedt}, {Kappelmann}, {Appenzeller},
  {Fromm}, {G{\"o}lz}, {Grewing}, {Gringel}, {Haas}, {Hopfensitz},
  {Kr{\"a}mer}, {Krautter}, {Lindenberger}, {Mandel}, \&
  {Widmann}}]{Barnstedt_etal_99}
{Barnstedt}, J., {Kappelmann}, N., {Appenzeller}, I., {et~al.} 1999, \aaps,
  134, 561

\bibitem[{{Blaauw}(1961)}]{Blaauw61}
{Blaauw}, A. 1961, \bain, 15, 265

\bibitem[{{Butler} \& {Giddings}(1985)}]{BuGi85}
{Butler}, K. \& {Giddings}, J.~R. 1985, Newsletter of Analysis of Astronomical
  Spectra, 9 (Univ. London)

\bibitem[{{Cantat-Gaudin} \& {Brandt}(2021)}]{CGB21}
{Cantat-Gaudin}, T. \& {Brandt}, T.~D. 2021, \aap, 649, A124

\bibitem[{{Cutri} {et~al.}(2003){Cutri}, {Skrutskie}, {van Dyk}, {Beichman},
  {Carpenter}, {Chester}, {Cambresy}, {Evans}, {Fowler}, {Gizis}, {Howard},
  {Huchra}, {Jarrett}, {Kopan}, {Kirkpatrick}, {Light}, {Marsh}, {McCallon},
  {Schneider}, {Stiening}, {Sykes}, {Weinberg}, {Wheaton}, {Wheelock}, \&
  {Zacarias}}]{Cutrietal03}
{Cutri}, R.~M., {Skrutskie}, M.~F., {van Dyk}, S., {et~al.} 2003, VizieR Online
  Data Catalog, II/246

\bibitem[{{Cutri} {et~al.}(2021){Cutri}, {Wright}, {Conrow}, {Fowler},
  {Eisenhardt}, {Grillmair}, {Kirkpatrick}, {Masci}, {McCallon}, {Wheelock},
  {Fajardo-Acosta}, {Yan}, {Benford}, {Harbut}, {Jarrett}, {Lake}, {Leisawitz},
  {Ressler}, {Stanford}, {Tsai}, {Liu}, {Helou}, {Mainzer}, {Gettngs},
  {Gonzalez}, {Hoffman}, {Marsh}, {Padgett}, {Skrutskie}, {Beck}, {Papin}, \&
  {Wittman}}]{Cutrietal21}
{Cutri}, R.~M., {Wright}, E.~L., {Conrow}, T., {et~al.} 2021, VizieR Online
  Data Catalog, II/328

\bibitem[{{Dekker} {et~al.}(2000){Dekker}, {D'Odorico}, {Kaufer}, {Delabre}, \&
  {Kotzlowski}}]{Dekkeretal00}
{Dekker}, H., {D'Odorico}, S., {Kaufer}, A., {Delabre}, B., \& {Kotzlowski}, H.
  2000, SPIE Conf. Ser., 4008, 534

\bibitem[{{Dixon} {et~al.}(2006){Dixon}, {Sankrit}, \& {Otte}}]{Dixonetal06}
{Dixon}, W. V.~D., {Sankrit}, R., \& {Otte}, B. 2006, \apj, 647, 328

\bibitem[{{Ekstr{\"o}m} {et~al.}(2012){Ekstr{\"o}m}, {Georgy}, {Eggenberger},
  {Meynet}, {Mowlavi}, {Wyttenbach}, {Granada}, {Decressin}, {Hirschi},
  {Frischknecht}, {Charbonnel}, \& {Maeder}}]{Ekstroemetal12}
{Ekstr{\"o}m}, S., {Georgy}, C., {Eggenberger}, P., {et~al.} 2012, \aap, 537,
  A146

\bibitem[{{El-Badry} {et~al.}(2021){El-Badry}, {Rix}, \&
  {Heintz}}]{ElBadryetal21}
{El-Badry}, K., {Rix}, H.-W., \& {Heintz}, T.~M. 2021, \mnras, 506, 2269

\bibitem[{{Feast} {et~al.}(1955){Feast}, {Thackeray}, \&
  {Wesselink}}]{Feastetal55}
{Feast}, M.~W., {Thackeray}, A.~D., \& {Wesselink}, A.~J. 1955, \memras, 67, 51

\bibitem[{{Filipovi{\'c}} {et~al.}(2022){Filipovi{\'c}}, {Payne}, {Alsaberi},
  {Norris}, {Macgregor}, {Rudnick}, {Koribalski}, {Leahy}, {Ducci}, {Kothes},
  {Andernach}, {Barnes}, {Boji{\v{c}}i{\'c}}, {Bozzetto}, {Brose}, {Collier},
  {Crawford}, {Crocker}, {Dai}, {Galvin}, {Haberl}, {Heber}, {Hill}, {Hopkins},
  {Hurley-Walker}, {Ingallinera}, {Jarrett}, {Kavanagh}, {Lenc}, {Luken},
  {Mackey}, {Manojlovi{\'c}}, {Maggi}, {Maitra}, {Pennock}, {Points}, {Riggi},
  {Rowell}, {Safi-Harb}, {Sano}, {Sasaki}, {Shabala}, {Stevens}, {van Loon},
  {Tothill}, {Umana}, {Uro{\v{s}}evi{\'c}}, {Velovi{\'c}}, {Vernstrom}, {West},
  \& {Wan}}]{Filipovicetal22}
{Filipovi{\'c}}, M.~D., {Payne}, J.~L., {Alsaberi}, R.~Z.~E., {et~al.} 2022,
  \mnras, 512, 265

\bibitem[{{Firnstein} \& {Przybilla}(2012)}]{FiPr12}
{Firnstein}, M. \& {Przybilla}, N. 2012, \aap, 543, A80

\bibitem[{{Fitzpatrick}(1999)}]{fitzpatrick99}
{Fitzpatrick}, E.~L. 1999, \pasp, 111, 63

\bibitem[{{Flynn} {et~al.}(2022){Flynn}, {Sekhri}, {Venville}, {Dixon},
  {Duffy}, {Mould}, \& {Taylor}}]{Flynnetal22}
{Flynn}, C., {Sekhri}, R., {Venville}, T., {et~al.} 2022, \mnras, 509, 4276

\bibitem[{{Fraser} {et~al.}(2010){Fraser}, {Dufton}, {Hunter}, \&
  {Ryans}}]{Fraser_etal_10}
{Fraser}, M., {Dufton}, P.~L., {Hunter}, I., \& {Ryans}, R.~S.~I. 2010, \mnras,
  404, 1306

\bibitem[{{Gaia Collaboration}(2022)}]{GaiaDR3}
{Gaia Collaboration}. 2022, VizieR Online Data Catalog, I/355

\bibitem[{{Gaia Collaboration} {et~al.}(2016){Gaia Collaboration}, {Prusti},
  {de Bruijne}, {Brown}, {Vallenari}, {Babusiaux}, {Bailer-Jones}, {Bastian},
  {Biermann}, {Evans}, {Eyer}, {Jansen}, {Jordi}, {Klioner}, {Lammers},
  {Lindegren}, {Luri}, {Mignard}, {Milligan}, {Panem}, {Poinsignon},
  {Pourbaix}, {Randich}, {Sarri}, {Sartoretti}, {Siddiqui}, {Soubiran},
  {Valette}, {van Leeuwen}, {Walton}, {Aerts}, {Arenou}, {Cropper}, {Drimmel},
  {H{\o}g}, {Katz}, {Lattanzi}, {O'Mullane}, {Grebel}, {Holland}, {Huc},
  {Passot}, {Bramante}, {Cacciari}, {Casta{\~n}eda}, {Chaoul}, {Cheek}, {De
  Angeli}, {Fabricius}, {Guerra}, {Hern{\'a}ndez}, {Jean-Antoine-Piccolo},
  {Masana}, {Messineo}, {Mowlavi}, {Nienartowicz}, {Ord{\'o}{\~n}ez-Blanco},
  {Panuzzo}, {Portell}, {Richards}, {Riello}, {Seabroke}, {Tanga},
  {Th{\'e}venin}, {Torra}, {Els}, {Gracia-Abril}, {Comoretto},
  {Garcia-Reinaldos}, {Lock}, {Mercier}, {Altmann}, {Andrae}, {Astraatmadja},
  {Bellas-Velidis}, {Benson}, {Berthier}, {Blomme}, {Busso}, {Carry},
  {Cellino}, {Clementini}, {Cowell}, {Creevey}, {Cuypers}, {Davidson}, {De
  Ridder}, {de Torres}, {Delchambre}, {Dell'Oro}, {Ducourant}, {Fr{\'e}mat},
  {Garc{\'\i}a-Torres}, {Gosset}, {Halbwachs}, {Hambly}, {Harrison}, {Hauser},
  {Hestroffer}, {Hodgkin}, {Huckle}, {Hutton}, {Jasniewicz}, {Jordan},
  {Kontizas}, {Korn}, {Lanzafame}, {Manteiga}, {Moitinho}, {Muinonen},
  {Osinde}, {Pancino}, {Pauwels}, {Petit}, {Recio-Blanco}, {Robin}, {Sarro},
  {Siopis}, {Smith}, {Smith}, {Sozzetti}, {Thuillot}, {van Reeven}, {Viala},
  {Abbas}, {Abreu Aramburu}, {Accart}, {Aguado}, {Allan}, {Allasia},
  {Altavilla}, {{\'A}lvarez}, {Alves}, {Anderson}, {Andrei}, {Anglada Varela},
  {Antiche}, {Antoja}, {Ant{\'o}n}, {Arcay}, {Atzei}, {Ayache}, {Bach},
  {Baker}, {Balaguer-N{\'u}{\~n}ez}, {Barache}, {Barata}, {Barbier}, {Barblan},
  {Baroni}, {Barrado y Navascu{\'e}s}, {Barros}, {Barstow}, {Becciani},
  {Bellazzini}, {Bellei}, {Bello Garc{\'\i}a}, {Belokurov}, {Bendjoya},
  {Berihuete}, {Bianchi}, {Bienaym{\'e}}, {Billebaud}, {Blagorodnova},
  {Blanco-Cuaresma}, {Boch}, {Bombrun}, {Borrachero}, {Bouquillon}, {Bourda},
  {Bouy}, {Bragaglia}, {Breddels}, {Brouillet}, {Br{\"u}semeister},
  {Bucciarelli}, {Budnik}, {Burgess}, {Burgon}, {Burlacu}, {Busonero}, {Buzzi},
  {Caffau}, {Cambras}, {Campbell}, {Cancelliere}, {Cantat-Gaudin}, {Carlucci},
  {Carrasco}, {Castellani}, {Charlot}, {Charnas}, {Charvet}, {Chassat},
  {Chiavassa}, {Clotet}, {Cocozza}, {Collins}, {Collins}, {Costigan}, {Crifo},
  {Cross}, {Crosta}, {Crowley}, {Dafonte}, {Damerdji}, {Dapergolas}, {David},
  {David}, {De Cat}, {de Felice}, {de Laverny}, {De Luise}, {De March}, {de
  Martino}, {de Souza}, {Debosscher}, {del Pozo}, {Delbo}, {Delgado},
  {Delgado}, {di Marco}, {Di Matteo}, {Diakite}, {Distefano}, {Dolding}, {Dos
  Anjos}, {Drazinos}, {Dur{\'a}n}, {Dzigan}, {Ecale}, {Edvardsson}, {Enke},
  {Erdmann}, {Escolar}, {Espina}, {Evans}, {Eynard Bontemps}, {Fabre},
  {Fabrizio}, {Faigler}, {Falc{\~a}o}, {Farr{\`a}s Casas}, {Faye}, {Federici},
  {Fedorets}, {Fern{\'a}ndez-Hern{\'a}ndez}, {Fernique}, {Fienga}, {Figueras},
  {Filippi}, {Findeisen}, {Fonti}, {Fouesneau}, {Fraile}, {Fraser}, {Fuchs},
  {Furnell}, {Gai}, {Galleti}, {Galluccio}, {Garabato}, {Garc{\'\i}a-Sedano},
  {Gar{\'e}}, {Garofalo}, {Garralda}, {Gavras}, {Gerssen}, {Geyer}, {Gilmore},
  {Girona}, {Giuffrida}, {Gomes}, {Gonz{\'a}lez-Marcos},
  {Gonz{\'a}lez-N{\'u}{\~n}ez}, {Gonz{\'a}lez-Vidal}, {Granvik}, {Guerrier},
  {Guillout}, {Guiraud}, {G{\'u}rpide}, {Guti{\'e}rrez-S{\'a}nchez}, {Guy},
  {Haigron}, {Hatzidimitriou}, {Haywood}, {Heiter}, {Helmi}, {Hobbs},
  {Hofmann}, {Holl}, {Holland}, {Hunt}, {Hypki}, {Icardi}, {Irwin}, {Jevardat
  de Fombelle}, {Jofr{\'e}}, {Jonker}, {Jorissen}, {Julbe}, {Karampelas},
  {Kochoska}, {Kohley}, {Kolenberg}, {Kontizas}, {Koposov}, {Kordopatis},
  {Koubsky}, {Kowalczyk}, {Krone-Martins}, {Kudryashova}, {Kull}, {Bachchan},
  {Lacoste-Seris}, {Lanza}, {Lavigne}, {Le Poncin-Lafitte}, {Lebreton},
  {Lebzelter}, {Leccia}, {Leclerc}, {Lecoeur-Taibi}, {Lemaitre}, {Lenhardt},
  {Leroux}, {Liao}, {Licata}, {Lindstr{\o}m}, {Lister}, {Livanou}, {Lobel},
  {L{\"o}ffler}, {L{\'o}pez}, {Lopez-Lozano}, {Lorenz}, {Loureiro},
  {MacDonald}, {Magalh{\~a}es Fernandes}, {Managau}, {Mann}, {Mantelet},
  {Marchal}, {Marchant}, {Marconi}, {Marie}, {Marinoni}, {Marrese},
  {Marschalk{\'o}}, {Marshall}, {Mart{\'\i}n-Fleitas}, {Martino}, {Mary},
  {Matijevi{\v{c}}}, {Mazeh}, {McMillan}, {Messina}, {Mestre}, {Michalik},
  {Millar}, {Miranda}, {Molina}, {Molinaro}, {Molinaro}, {Moln{\'a}r},
  {Moniez}, {Montegriffo}, {Monteiro}, {Mor}, {Mora}, {Morbidelli}, {Morel},
  {Morgenthaler}, {Morley}, {Morris}, {Mulone}, {Muraveva}, {Musella},
  {Narbonne}, {Nelemans}, {Nicastro}, {Noval}, {Ord{\'e}novic},
  {Ordieres-Mer{\'e}}, {Osborne}, {Pagani}, {Pagano}, {Pailler}, {Palacin},
  {Palaversa}, {Parsons}, {Paulsen}, {Pecoraro}, {Pedrosa}, {Pentik{\"a}inen},
  {Pereira}, {Pichon}, {Piersimoni}, {Pineau}, {Plachy}, {Plum}, {Poujoulet},
  {Pr{\v{s}}a}, {Pulone}, {Ragaini}, {Rago}, {Rambaux}, {Ramos-Lerate},
  {Ranalli}, {Rauw}, {Read}, {Regibo}, {Renk}, {Reyl{\'e}}, {Ribeiro},
  {Rimoldini}, {Ripepi}, {Riva}, {Rixon}, {Roelens}, {Romero-G{\'o}mez},
  {Rowell}, {Royer}, {Rudolph}, {Ruiz-Dern}, {Sadowski}, {Sagrist{\`a}
  Sell{\'e}s}, {Sahlmann}, {Salgado}, {Salguero}, {Sarasso}, {Savietto},
  {Schnorhk}, {Schultheis}, {Sciacca}, {Segol}, {Segovia}, {Segransan},
  {Serpell}, {Shih}, {Smareglia}, {Smart}, {Smith}, {Solano}, {Solitro},
  {Sordo}, {Soria Nieto}, {Souchay}, {Spagna}, {Spoto}, {Stampa}, {Steele},
  {Steidelm{\"u}ller}, {Stephenson}, {Stoev}, {Suess}, {S{\"u}veges}, {Surdej},
  {Szabados}, {Szegedi-Elek}, {Tapiador}, {Taris}, {Tauran}, {Taylor},
  {Teixeira}, {Terrett}, {Tingley}, {Trager}, {Turon}, {Ulla}, {Utrilla},
  {Valentini}, {van Elteren}, {Van Hemelryck}, {van Leeuwen}, {Varadi},
  {Vecchiato}, {Veljanoski}, {Via}, {Vicente}, {Vogt}, {Voss}, {Votruba},
  {Voutsinas}, {Walmsley}, {Weiler}, {Weingrill}, {Werner}, {Wevers},
  {Whitehead}, {Wyrzykowski}, {Yoldas}, {{\v{Z}}erjal}, {Zucker}, {Zurbach},
  {Zwitter}, {Alecu}, {Allen}, {Allende Prieto}, {Amorim},
  {Anglada-Escud{\'e}}, {Arsenijevic}, {Azaz}, {Balm}, {Beck}, {Bernstein},
  {Bigot}, {Bijaoui}, {Blasco}, {Bonfigli}, {Bono}, {Boudreault}, {Bressan},
  {Brown}, {Brunet}, {Bunclark}, {Buonanno}, {Butkevich}, {Carret}, {Carrion},
  {Chemin}, {Ch{\'e}reau}, {Corcione}, {Darmigny}, {de Boer}, {de Teodoro}, {de
  Zeeuw}, {Delle Luche}, {Domingues}, {Dubath}, {Fodor}, {Fr{\'e}zouls},
  {Fries}, {Fustes}, {Fyfe}, {Gallardo}, {Gallegos}, {Gardiol}, {Gebran},
  {Gomboc}, {G{\'o}mez}, {Grux}, {Gueguen}, {Heyrovsky}, {Hoar}, {Iannicola},
  {Isasi Parache}, {Janotto}, {Joliet}, {Jonckheere}, {Keil}, {Kim},
  {Klagyivik}, {Klar}, {Knude}, {Kochukhov}, {Kolka}, {Kos}, {Kutka}, {Lainey},
  {LeBouquin}, {Liu}, {Loreggia}, {Makarov}, {Marseille}, {Martayan},
  {Martinez-Rubi}, {Massart}, {Meynadier}, {Mignot}, {Munari}, {Nguyen},
  {Nordlander}, {Ocvirk}, {O'Flaherty}, {Olias Sanz}, {Ortiz}, {Osorio},
  {Oszkiewicz}, {Ouzounis}, {Palmer}, {Park}, {Pasquato}, {Peltzer}, {Peralta},
  {P{\'e}turaud}, {Pieniluoma}, {Pigozzi}, {Poels}, {Prat}, {Prod'homme},
  {Raison}, {Rebordao}, {Risquez}, {Rocca-Volmerange}, {Rosen}, {Ruiz-Fuertes},
  {Russo}, {Sembay}, {Serraller Vizcaino}, {Short}, {Siebert}, {Silva},
  {Sinachopoulos}, {Slezak}, {Soffel}, {Sosnowska}, {Strai{\v{z}}ys}, {ter
  Linden}, {Terrell}, {Theil}, {Tiede}, {Troisi}, {Tsalmantza}, {Tur},
  {Vaccari}, {Vachier}, {Valles}, {Van Hamme}, {Veltz}, {Virtanen}, {Wallut},
  {Wichmann}, {Wilkinson}, {Ziaeepour}, \& {Zschocke}}]{Gaia2016}
{Gaia Collaboration}, {Prusti}, T., {de Bruijne}, J.~H.~J., {et~al.} 2016,
  \aap, 595, A1

\bibitem[{{Gaia Collaboration} {et~al.}(2023){Gaia Collaboration}, {Vallenari},
  {Brown}, {Prusti}, {de Bruijne}, {Arenou}, {Babusiaux}, {Biermann},
  {Creevey}, {Ducourant}, {Evans}, {Eyer}, {Guerra}, {Hutton}, {Jordi},
  {Klioner}, {Lammers}, {Lindegren}, {Luri}, {Mignard}, {Panem}, {Pourbaix},
  {Randich}, {Sartoretti}, {Soubiran}, {Tanga}, {Walton}, {Bailer-Jones},
  {Bastian}, {Drimmel}, {Jansen}, {Katz}, {Lattanzi}, {van Leeuwen}, {Bakker},
  {Cacciari}, {Casta{\~n}eda}, {De Angeli}, {Fabricius}, {Fouesneau},
  {Fr{\'e}mat}, {Galluccio}, {Guerrier}, {Heiter}, {Masana}, {Messineo},
  {Mowlavi}, {Nicolas}, {Nienartowicz}, {Pailler}, {Panuzzo}, {Riclet}, {Roux},
  {Seabroke}, {Sordo}, {Th{\'e}venin}, {Gracia-Abril}, {Portell}, {Teyssier},
  {Altmann}, {Andrae}, {Audard}, {Bellas-Velidis}, {Benson}, {Berthier},
  {Blomme}, {Burgess}, {Busonero}, {Busso}, {C{\'a}novas}, {Carry}, {Cellino},
  {Cheek}, {Clementini}, {Damerdji}, {Davidson}, {de Teodoro}, {Nu{\~n}ez
  Campos}, {Delchambre}, {Dell'Oro}, {Esquej}, {Fern{\'a}ndez-Hern{\'a}ndez},
  {Fraile}, {Garabato}, {Garc{\'\i}a-Lario}, {Gosset}, {Haigron}, {Halbwachs},
  {Hambly}, {Harrison}, {Hern{\'a}ndez}, {Hestroffer}, {Hodgkin}, {Holl},
  {Jan{\ss}en}, {Jevardat de Fombelle}, {Jordan}, {Krone-Martins}, {Lanzafame},
  {L{\"o}ffler}, {Marchal}, {Marrese}, {Moitinho}, {Muinonen}, {Osborne},
  {Pancino}, {Pauwels}, {Recio-Blanco}, {Reyl{\'e}}, {Riello}, {Rimoldini},
  {Roegiers}, {Rybizki}, {Sarro}, {Siopis}, {Smith}, {Sozzetti}, {Utrilla},
  {van Leeuwen}, {Abbas}, {{\'A}brah{\'a}m}, {Abreu Aramburu}, {Aerts},
  {Aguado}, {Ajaj}, {Aldea-Montero}, {Altavilla}, {{\'A}lvarez}, {Alves},
  {Anders}, {Anderson}, {Anglada Varela}, {Antoja}, {Baines}, {Baker},
  {Balaguer-N{\'u}{\~n}ez}, {Balbinot}, {Balog}, {Barache}, {Barbato},
  {Barros}, {Barstow}, {Bartolom{\'e}}, {Bassilana}, {Bauchet}, {Becciani},
  {Bellazzini}, {Berihuete}, {Bernet}, {Bertone}, {Bianchi}, {Binnenfeld},
  {Blanco-Cuaresma}, {Blazere}, {Boch}, {Bombrun}, {Bossini}, {Bouquillon},
  {Bragaglia}, {Bramante}, {Breedt}, {Bressan}, {Brouillet}, {Brugaletta},
  {Bucciarelli}, {Burlacu}, {Butkevich}, {Buzzi}, {Caffau}, {Cancelliere},
  {Cantat-Gaudin}, {Carballo}, {Carlucci}, {Carnerero}, {Carrasco},
  {Casamiquela}, {Castellani}, {Castro-Ginard}, {Chaoul}, {Charlot}, {Chemin},
  {Chiaramida}, {Chiavassa}, {Chornay}, {Comoretto}, {Contursi}, {Cooper},
  {Cornez}, {Cowell}, {Crifo}, {Cropper}, {Crosta}, {Crowley}, {Dafonte},
  {Dapergolas}, {David}, {David}, {de Laverny}, {De Luise}, {De March}, {De
  Ridder}, {de Souza}, {de Torres}, {del Peloso}, {del Pozo}, {Delbo},
  {Delgado}, {Delisle}, {Demouchy}, {Dharmawardena}, {Di Matteo}, {Diakite},
  {Diener}, {Distefano}, {Dolding}, {Edvardsson}, {Enke}, {Fabre}, {Fabrizio},
  {Faigler}, {Fedorets}, {Fernique}, {Fienga}, {Figueras}, {Fournier},
  {Fouron}, {Fragkoudi}, {Gai}, {Garcia-Gutierrez}, {Garcia-Reinaldos},
  {Garc{\'\i}a-Torres}, {Garofalo}, {Gavel}, {Gavras}, {Gerlach}, {Geyer},
  {Giacobbe}, {Gilmore}, {Girona}, {Giuffrida}, {Gomel}, {Gomez},
  {Gonz{\'a}lez-N{\'u}{\~n}ez}, {Gonz{\'a}lez-Santamar{\'\i}a},
  {Gonz{\'a}lez-Vidal}, {Granvik}, {Guillout}, {Guiraud},
  {Guti{\'e}rrez-S{\'a}nchez}, {Guy}, {Hatzidimitriou}, {Hauser}, {Haywood},
  {Helmer}, {Helmi}, {Sarmiento}, {Hidalgo}, {Hilger}, {H{\l}adczuk}, {Hobbs},
  {Holland}, {Huckle}, {Jardine}, {Jasniewicz}, {Jean-Antoine Piccolo},
  {Jim{\'e}nez-Arranz}, {Jorissen}, {Juaristi Campillo}, {Julbe}, {Karbevska},
  {Kervella}, {Khanna}, {Kontizas}, {Kordopatis}, {Korn}, {K{\'o}sp{\'a}l},
  {Kostrzewa-Rutkowska}, {Kruszy{\'n}ska}, {Kun}, {Laizeau}, {Lambert},
  {Lanza}, {Lasne}, {Le Campion}, {Lebreton}, {Lebzelter}, {Leccia}, {Leclerc},
  {Lecoeur-Taibi}, {Liao}, {Licata}, {Lindstr{\o}m}, {Lister}, {Livanou},
  {Lobel}, {Lorca}, {Loup}, {Madrero Pardo}, {Magdaleno Romeo}, {Managau},
  {Mann}, {Manteiga}, {Marchant}, {Marconi}, {Marcos}, {Marcos Santos},
  {Mar{\'\i}n Pina}, {Marinoni}, {Marocco}, {Marshall}, {Martin Polo},
  {Mart{\'\i}n-Fleitas}, {Marton}, {Mary}, {Masip}, {Massari},
  {Mastrobuono-Battisti}, {Mazeh}, {McMillan}, {Messina}, {Michalik}, {Millar},
  {Mints}, {Molina}, {Molinaro}, {Moln{\'a}r}, {Monari}, {Mongui{\'o}},
  {Montegriffo}, {Montero}, {Mor}, {Mora}, {Morbidelli}, {Morel}, {Morris},
  {Muraveva}, {Murphy}, {Musella}, {Nagy}, {Noval}, {Oca{\~n}a}, {Ogden},
  {Ordenovic}, {Osinde}, {Pagani}, {Pagano}, {Palaversa}, {Palicio},
  {Pallas-Quintela}, {Panahi}, {Payne-Wardenaar}, {Pe{\~n}alosa Esteller},
  {Penttil{\"a}}, {Pichon}, {Piersimoni}, {Pineau}, {Plachy}, {Plum}, {Poggio},
  {Pr{\v{s}}a}, {Pulone}, {Racero}, {Ragaini}, {Rainer}, {Raiteri}, {Rambaux},
  {Ramos}, {Ramos-Lerate}, {Re Fiorentin}, {Regibo}, {Richards}, {Rios Diaz},
  {Ripepi}, {Riva}, {Rix}, {Rixon}, {Robichon}, {Robin}, {Robin}, {Roelens},
  {Rogues}, {Rohrbasser}, {Romero-G{\'o}mez}, {Rowell}, {Royer}, {Ruz Mieres},
  {Rybicki}, {Sadowski}, {S{\'a}ez N{\'u}{\~n}ez}, {Sagrist{\`a} Sell{\'e}s},
  {Sahlmann}, {Salguero}, {Samaras}, {Sanchez Gimenez}, {Sanna},
  {Santove{\~n}a}, {Sarasso}, {Schultheis}, {Sciacca}, {Segol}, {Segovia},
  {S{\'e}gransan}, {Semeux}, {Shahaf}, {Siddiqui}, {Siebert}, {Siltala},
  {Silvelo}, {Slezak}, {Slezak}, {Smart}, {Snaith}, {Solano}, {Solitro},
  {Souami}, {Souchay}, {Spagna}, {Spina}, {Spoto}, {Steele},
  {Steidelm{\"u}ller}, {Stephenson}, {S{\"u}veges}, {Surdej}, {Szabados},
  {Szegedi-Elek}, {Taris}, {Taylor}, {Teixeira}, {Tolomei}, {Tonello}, {Torra},
  {Torra}, {Torralba Elipe}, {Trabucchi}, {Tsounis}, {Turon}, {Ulla}, {Unger},
  {Vaillant}, {van Dillen}, {van Reeven}, {Vanel}, {Vecchiato}, {Viala},
  {Vicente}, {Voutsinas}, {Weiler}, {Wevers}, {Wyrzykowski}, {Yoldas}, {Yvard},
  {Zhao}, {Zorec}, {Zucker}, \& {Zwitter}}]{Gaia2023}
{Gaia Collaboration}, {Vallenari}, A., {Brown}, A.~G.~A., {et~al.} 2023, \aap,
  674, A1

\bibitem[{{Georgy} {et~al.}(2013){Georgy}, {Ekstr{\"o}m}, {Eggenberger},
  {Meynet}, {Haemmerl{\'e}}, {Maeder}, {Granada}, {Groh}, {Hirschi}, {Mowlavi},
  {Yusof}, {Charbonnel}, {Decressin}, \& {Barblan}}]{Georgyetal13}
{Georgy}, C., {Ekstr{\"o}m}, S., {Eggenberger}, P., {et~al.} 2013, \aap, 558,
  A103

\bibitem[{{Giddings}(1981)}]{Giddings81}
{Giddings}, J.~R. 1981, PhD thesis, (Univ. London)

\bibitem[{{Gies} \& {Bolton}(1986)}]{GiBo86}
{Gies}, D.~R. \& {Bolton}, C.~T. 1986, \apjs, 61, 419

\bibitem[{{Gravity Collaboration} {et~al.}(2019){Gravity Collaboration},
  {Abuter}, {Amorim}, {Baub{\"o}ck}, {Berger}, {Bonnet}, {Brandner},
  {Cl{\'e}net}, {Coud{\'e} Du Foresto}, {de Zeeuw}, {Dexter}, {Duvert},
  {Eckart}, {Eisenhauer}, {F{\"o}rster Schreiber}, {Garcia}, {Gao}, {Gendron},
  {Genzel}, {Gerhard}, {Gillessen}, {Habibi}, {Haubois}, {Henning}, {Hippler},
  {Horrobin}, {Jim{\'e}nez-Rosales}, {Jocou}, {Kervella}, {Lacour},
  {Lapeyr{\`e}re}, {Le Bouquin}, {L{\'e}na}, {Ott}, {Paumard}, {Perraut},
  {Perrin}, {Pfuhl}, {Rabien}, {Rodriguez Coira}, {Rousset}, {Scheithauer},
  {Sternberg}, {Straub}, {Straubmeier}, {Sturm}, {Tacconi}, {Vincent}, {von
  Fellenberg}, {Waisberg}, {Widmann}, {Wieprecht}, {Wiezorrek}, {Woillez}, \&
  {Yazici}}]{GravityCol1aboration19}
{Gravity Collaboration}, {Abuter}, R., {Amorim}, A., {et~al.} 2019, \aap, 625,
  L10

\bibitem[{{Herrero} {et~al.}(2022){Herrero}, {Berlanas}, {Gil de Paz},
  {Comer{\'o}n}, {Puls}, {Ram{\'\i}rez Alegr{\'\i}a}, {Garcia}, {Lennon},
  {Najarro}, {Sim{\'o}n-D{\'\i}az}, {Urbaneja}, {Gallego}, {Carrasco},
  {Iglesias}, {Cedazo}, {Garc{\'\i}a Vargas}, {Castillo-Morales}, {Pascual},
  {Cardiel}, {P{\'e}rez-Calpena}, {G{\'o}mez-Alvarez}, \&
  {Mart{\'\i}nez-Delgado}}]{Herreroetal22}
{Herrero}, A., {Berlanas}, S.~R., {Gil de Paz}, A., {et~al.} 2022, \mnras, 511,
  3113

\bibitem[{{Hirsch}(2009)}]{Hirsch09}
{Hirsch}, H.~A. 2009, PhD thesis, (Univ. Erlangen-N{\"u}rnberg)

\bibitem[{{Irrgang} {et~al.}(2010){Irrgang}, {Przybilla}, {Heber}, {Nieva}, \&
  {Schuh}}]{Irrgangetal10}
{Irrgang}, A., {Przybilla}, N., {Heber}, U., {Nieva}, M.~F., \& {Schuh}, S.
  2010, \apj, 711, 138

\bibitem[{{Irrgang} {et~al.}(2022){Irrgang}, {Przybilla}, \&
  {Meynet}}]{Irrgangetal22}
{Irrgang}, A., {Przybilla}, N., \& {Meynet}, G. 2022, Nat. Astron., 6, 1414

\bibitem[{{Kaufer} {et~al.}(1999){Kaufer}, {Stahl}, {Tubbesing},
  {N{\o}rregaard}, {Avila}, {Francois}, {Pasquini}, \&
  {Pizzella}}]{Kauferetal99}
{Kaufer}, A., {Stahl}, O., {Tubbesing}, S., {et~al.} 1999, The Messenger, 95, 8

\bibitem[{{Kippenhahn} {et~al.}(2013){Kippenhahn}, {Weigert}, \&
  {Weiss}}]{Kippenhahnetal13}
{Kippenhahn}, R., {Weigert}, A., \& {Weiss}, A. 2013, {Stellar Structure and
  Evolution, 2nd ed.} (Springer, Berlin Heidelberg)

\bibitem[{{Kudritzki} {et~al.}(2016){Kudritzki}, {Castro}, {Urbaneja}, {Ho},
  {Bresolin}, {Gieren}, {Pietrzy{\'n}ski}, \& {Przybilla}}]{Kudritzkietal16}
{Kudritzki}, R.~P., {Castro}, N., {Urbaneja}, M.~A., {et~al.} 2016, \apj, 829,
  70

\bibitem[{{Kurucz}(2005)}]{kurucz05}
{Kurucz}, R.~L. 2005, Mem. Soc. Astron. Ital. Suppl., 8, 14

\bibitem[{{Lindegren} {et~al.}(2021){Lindegren}, {Bastian}, {Biermann},
  {Bombrun}, {de Torres}, {Gerlach}, {Geyer}, {Hern{\'a}ndez}, {Hilger},
  {Hobbs}, {Klioner}, {Lammers}, {McMillan}, {Ramos-Lerate},
  {Steidelm{\"u}ller}, {Stephenson}, \& {van Leeuwen}}]{Lindegrenetal21}
{Lindegren}, L., {Bastian}, U., {Biermann}, M., {et~al.} 2021, \aap, 649, A4

\bibitem[{{Mackereth} {et~al.}(2017){Mackereth}, {Bovy}, {Schiavon},
  {Zasowski}, {Cunha}, {Frinchaboy}, {Garc{\'\i}a Perez}, {Hayden}, {Holtzman},
  {Majewski}, {M{\'e}sz{\'a}ros}, {Nidever}, {Pinsonneault}, \&
  {Shetrone}}]{Mackerethetal17}
{Mackereth}, J.~T., {Bovy}, J., {Schiavon}, R.~P., {et~al.} 2017, \mnras, 471,
  3057

\bibitem[{{Maeder} {et~al.}(2014){Maeder}, {Przybilla}, {Nieva}, {Georgy},
  {Meynet}, {Ekstr{\"o}m}, \& {Eggenberger}}]{Maederetal14}
{Maeder}, A., {Przybilla}, N., {Nieva}, M.~F., {et~al.} 2014, \aap, 565, A39

\bibitem[{{Ma{\'\i}z-Apell{\'a}niz}(2001)}]{MaizApellaniz01}
{Ma{\'\i}z-Apell{\'a}niz}, J. 2001, \aj, 121, 2737

\bibitem[{{Ma{\'\i}z Apell{\'a}niz}(2022)}]{MaizApellaniz22}
{Ma{\'\i}z Apell{\'a}niz}, J. 2022, \aap, 657, A130

\bibitem[{{Ma{\'\i}z Apell{\'a}niz} {et~al.}(2008){Ma{\'\i}z Apell{\'a}niz},
  {Alfaro}, \& {Sota}}]{MaizApellanizetal2008}
{Ma{\'\i}z Apell{\'a}niz}, J., {Alfaro}, E.~J., \& {Sota}, A. 2008, arXiv
  e-prints, arXiv:0804.2553

\bibitem[{{Massa} {et~al.}(1991){Massa}, {Altner}, {Wynne}, \&
  {Lamers}}]{Massaetal91}
{Massa}, D., {Altner}, B., {Wynne}, D., \& {Lamers}, H. J. G. L.~M. 1991, \aap,
  242, 188

\bibitem[{{Mermilliod}(1997)}]{Mermilliod97}
{Mermilliod}, J.~C. 1997, VizieR Online Data Catalog, 2168

\bibitem[{{Nieva} \& {Przybilla}(2006)}]{NiPr06}
{Nieva}, M.~F. \& {Przybilla}, N. 2006, ApJ, 639, L39

\bibitem[{{Nieva} \& {Przybilla}(2007)}]{NiPr07}
{Nieva}, M.~F. \& {Przybilla}, N. 2007, \aap, 467, 295

\bibitem[{{Nieva} \& {Przybilla}(2012)}]{NiPr12}
{Nieva}, M.~F. \& {Przybilla}, N. 2012, \aap, 539, A143

\bibitem[{{Nieva} \& {Sim{\'o}n-D{\'\i}az}(2011)}]{NiSi11}
{Nieva}, M.~F. \& {Sim{\'o}n-D{\'\i}az}, S. 2011, \aap, 532, A2

\bibitem[{{Normandeau} {et~al.}(1996){Normandeau}, {Taylor}, \&
  {Dewdney}}]{Normandeauetal96}
{Normandeau}, M., {Taylor}, A.~R., \& {Dewdney}, P.~E. 1996, \nat, 380, 687

\bibitem[{{Odenkirchen} \& {Brosche}(1992)}]{OdBr92}
{Odenkirchen}, M. \& {Brosche}, P. 1992, Astronomische Nachrichten, 313, 69

\bibitem[{{Pantaleoni Gonz{\'a}lez} {et~al.}(2021){Pantaleoni Gonz{\'a}lez},
  {Reed}}]{PantaleoniGonzalezetal21}
{Pantaleoni Gonz{\'a}lez}, M., 
  R.~H., \& {Reed}, B.~C. 2021, \mnras, 504, 2968

\bibitem[{{Podsiadlowski}(1992)}]{Podsiadlowski92}
{Podsiadlowski}, P. 1992, \pasp, 104, 717

\bibitem[{{Prinja} \& {Massa}(2013)}]{Prinjaetal2013}
{Prinja}, R.~K. \& {Massa}, D.~L. 2013, \aap, 559, A15

\bibitem[{{Przybilla} \& {Butler}(2001)}]{PrBu01}
{Przybilla}, N. \& {Butler}, K. 2001, \aap, 379, 955

\bibitem[{{Przybilla} {et~al.}(2006){Przybilla}, {Butler}, {Becker}, \&
  {Kudritzki}}]{Przybillaetal06}
{Przybilla}, N., {Butler}, K., {Becker}, S.~R., \& {Kudritzki}, R.~P. 2006,
  \aap, 445, 1099

\bibitem[{{Przybilla} {et~al.}(2001){Przybilla}, {Butler}, \&
  {Kudritzki}}]{Przybillaetal01b}
{Przybilla}, N., {Butler}, K., \& {Kudritzki}, R.~P. 2001, \aap, 379, 936

\bibitem[{{Przybilla} {et~al.}(2010){Przybilla}, {Firnstein}, {Nieva},
  {Meynet}, \& {Maeder}}]{Przybillaetal10}
{Przybilla}, N., {Firnstein}, M., {Nieva}, M.~F., {Meynet}, G., \& {Maeder}, A.
  2010, \aap, 517, A38

\bibitem[{{Przybilla} {et~al.}(2008{\natexlab{a}}){Przybilla}, {Nieva}, \&
  {Butler}}]{Przybillaetal08b}
{Przybilla}, N., {Nieva}, M.~F., \& {Butler}, K. 2008{\natexlab{a}}, \apjl,
  688, L103

\bibitem[{{Przybilla} {et~al.}(2008{\natexlab{b}}){Przybilla}, {Nieva},
  {Heber}, \& {Butler}}]{Przybillaetal08c}
{Przybilla}, N., {Nieva}, M.~F., {Heber}, U., \& {Butler}, K.
  2008{\natexlab{b}}, \apjl, 684, L103

\bibitem[{{Przybilla} {et~al.}(2008{\natexlab{c}}){Przybilla}, {Nieva},
  {Heber}, {Firnstein}, {Butler}, {Napiwotzki}, \&
  {Edelmann}}]{Przybillaetal08d}
{Przybilla}, N., {Nieva}, M.~F., {Heber}, U., {et~al.} 2008{\natexlab{c}},
  \aap, 480, L37

\bibitem[{{Puls} {et~al.}(2020){Puls}, {Najarro}, {Sundqvist}, \&
  {Sen}}]{Pulsetal20}
{Puls}, J., {Najarro}, F., {Sundqvist}, J.~O., \& {Sen}, K. 2020, \aap, 642,
  A172

\bibitem[{{Puls} {et~al.}(2005){Puls}, {Urbaneja}, {Venero}, {Repolust},
  {Springmann}, {Jokuthy}, \& {Mokiem}}]{Pulsetal05}
{Puls}, J., {Urbaneja}, M.~A., {Venero}, R., {et~al.} 2005, \aap, 435, 669

\bibitem[{{Savage} \& {Massa}(1987)}]{SaMa87}
{Savage}, B.~D. \& {Massa}, D. 1987, \apj, 314, 380

\bibitem[{{Sen} {et~al.}(2022){Sen}, {Langer}, {Marchant}, {Menon}, {de Mink},
  {Schootemeijer}, {Sch{\"u}rmann}, {Mahy}, {Hastings}, {Nathaniel}, {Sana},
  {Wang}, \& {Xu}}]{Senetal22}
{Sen}, K., {Langer}, N., {Marchant}, P., {et~al.} 2022, \aap, 659, A98

\bibitem[{{Silva} \& {Napiwotzki}(2011)}]{SiNa11}
{Silva}, M.~D.~V. \& {Napiwotzki}, R. 2011, \mnras, 411, 2596

\bibitem[{{Smartt}(2009)}]{Smartt09}
{Smartt}, S.~J. 2009, \araa, 47, 63

\bibitem[{{Tenorio-Tagle}(1996)}]{Tenorio-Tagle96}
{Tenorio-Tagle}, G. 1996, \aj, 111, 1641

\bibitem[{{Thompson} {et~al.}(1995){Thompson}, {Nandy}, {Jamar}, {Monfils},
  {Houziaux L.}, {Carnochan}, \& {Wilson}}]{Thompson95}
{Thompson}, G.~I., {Nandy}, K., {Jamar}, C., {et~al.} 1995, VizieR Online Data
  Catalog, II/59B

\bibitem[{{Urbaneja} {et~al.}(2005){Urbaneja}, {Herrero}, {Bresolin},
  {Kudritzki}, {Gieren}, {Puls}, {Przybilla}, {Najarro}, \&
  {Pietrzy{\'n}ski}}]{Urbanejaetal05a}
{Urbaneja}, M.~A., {Herrero}, A., {Bresolin}, F., {et~al.} 2005, \apj, 622, 862

\bibitem[{{Urbaneja} {et~al.}(2017){Urbaneja}, {Kudritzki}, {Gieren},
  {Pietrzy{\'n}ski}, {Bresolin}, \& {Przybilla}}]{Urbanejaetal17}
{Urbaneja}, M.~A., {Kudritzki}, R.~P., {Gieren}, W., {et~al.} 2017, \aj, 154,
  102

\bibitem[{{Vrancken} {et~al.}(1996){Vrancken}, {Butler}, \&
  {Becker}}]{Vranckenetal96}
{Vrancken}, M., {Butler}, K., \& {Becker}, S.~R. 1996, \aap, 311, 661

\bibitem[{{Walborn} {et~al.}(1990){Walborn}, {Fitzpatrick}, \&
  {Nichols-Bohlin}}]{Walbornetal90}
{Walborn}, N.~R., {Fitzpatrick}, E.~L., \& {Nichols-Bohlin}, J. 1990, \pasp,
  102, 543

\bibitem[{{Walborn} {et~al.}(1989){Walborn}, {Prevot}, {Prevot}, {Wamsteker},
  {Gonzalez}, {Gilmozzi}, \& {Fitzpatrick}}]{Walbornetal89}
{Walborn}, N.~R., {Prevot}, M.~L., {Prevot}, L., {et~al.} 1989, \aap, 219, 229

\bibitem[{{Wang} {et~al.}(2021){Wang}, {Gies}, {Peters}, {G{\"o}tberg},
  {Chojnowski}, {Lester}, \& {Howell}}]{Wangetal21}
{Wang}, L., {Gies}, D.~R., {Peters}, G.~J., {et~al.} 2021, \aj, 161, 248

\bibitem[{{Wellstein} {et~al.}(2001){Wellstein}, {Langer}, \&
  {Braun}}]{Wellsteinetal01}
{Wellstein}, S., {Langer}, N., \& {Braun}, H. 2001, \aap, 369, 939

\bibitem[{{We{\ss}mayer} {et~al.}(2022){We{\ss}mayer}, {Przybilla}, \&
  {Butler}}]{Wessmayeretal22}
{We{\ss}mayer}, D., {Przybilla}, N., \& {Butler}, K. 2022, \aap, 668, A92

\bibitem[{{We{\ss}mayer} {et~al.}(2023){We{\ss}mayer}, {Przybilla},
  {Ebenbichler}, {Aschenbrenner}, \& {Butler}}]{Wessmayeretal23}
{We{\ss}mayer}, D., {Przybilla}, N., {Ebenbichler}, A., {Aschenbrenner}, P., \&
  {Butler}, K. 2023, \aap, 677, A175

\bibitem[{{Widmann} {et~al.}(1998){Widmann}, {de Boer}, {Richter}, {Kramer},
  {Appenzeller}, {Barnstedt}, {Golz}, {Grewing}, {Gringel}, {Mandel}, \&
  {Werner}}]{Widmannetal98}
{Widmann}, H., {de Boer}, K.~S., {Richter}, P., {et~al.} 1998, \aap, 338, L1

\end{thebibliography}

\begin{appendix} 

\section{Spectral analysis of HD~93840}\label{appendix:A}
The determination of the atmospheric parameters from the high-resolution optical spectrum proceeded as follows. Effective temperature, $T_\mathrm{eff}$, and surface gravity, $\log g$, were established by simultaneously reproducing the Stark-broadened higher Balmer lines and ionisation equilibria of \ion{He}{i/ii} and multiple metallic species (e.g. \ion{O}{\sc i/ii} and \ion{Si}{\sc ii/iii/iv}). The microturbulent velocity, $\xi$, was constrained by requiring elemental abundances to  be independent of line equivalent width. Fitting individual line profiles then established the projected rotational velocity, $\varv \sin i$, and the macroturbulent velocity , $\zeta$, (applying a convolution with rotational and radial-tangential profiles), and the elemental abundances,~$\varepsilon(X)$\,=\,$\log(X/$H)\,+\,12. The metallicity, $Z$, was computed from the abundances of the elements discussed here, which cover the most important chemical species.
 
\begin{figure*}
\centering 
\includegraphics[angle=90,width=.98\hsize]{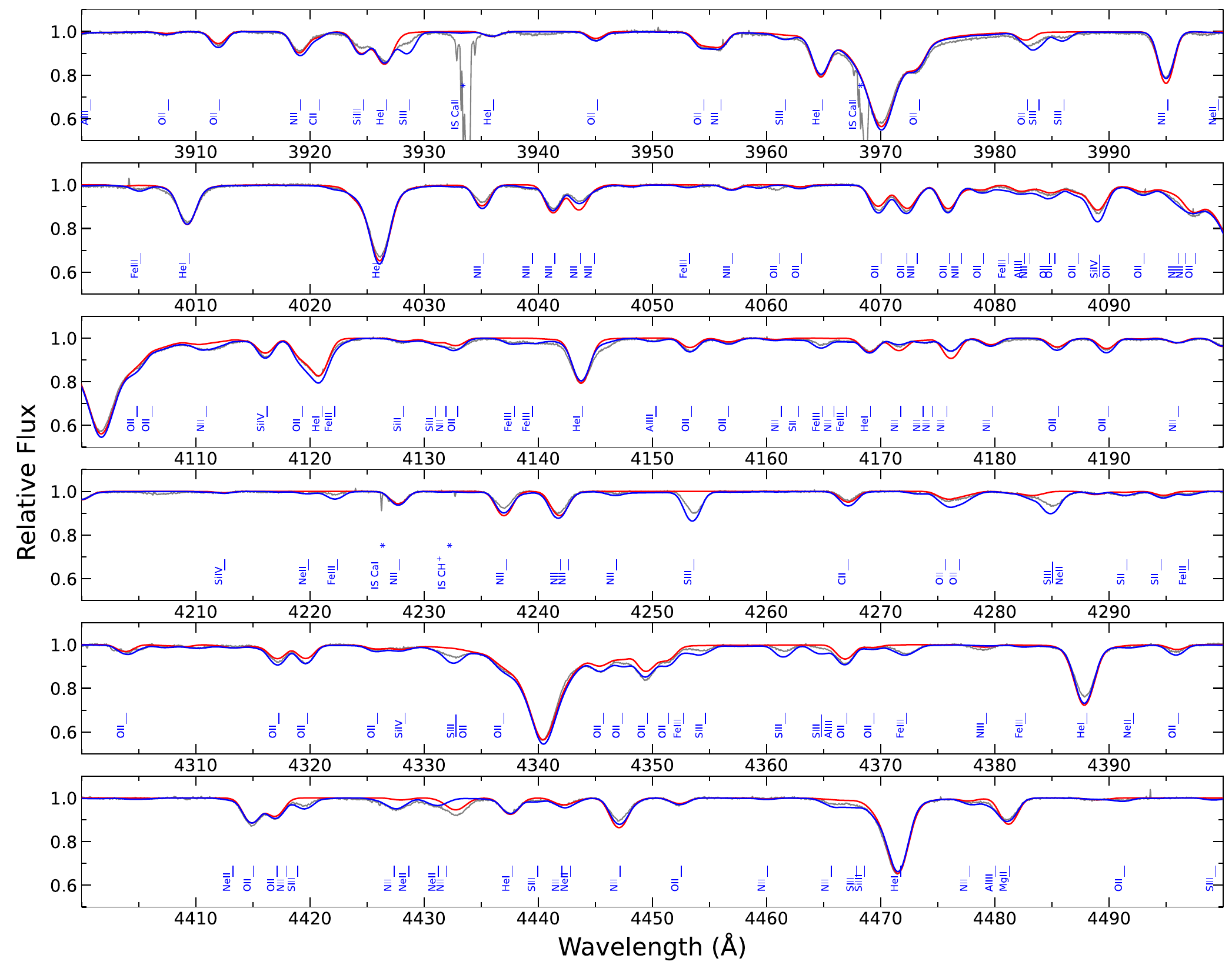}
        \caption{Comparison between the observed spectrum of HD~93840 (grey) and the best fitting synthetic spectra calculated with {\sc Ads} (blue) and {\sc fastwind} (red) in the wavelength range of 3900 to 4500\,{\AA}. The observed stellar spectrum was shifted to the laboratory rest frame.}
    \label{fig:hd93840_1}
\end{figure*}

\begin{figure*}
\centering 
\includegraphics[angle=90,width=.98\hsize]{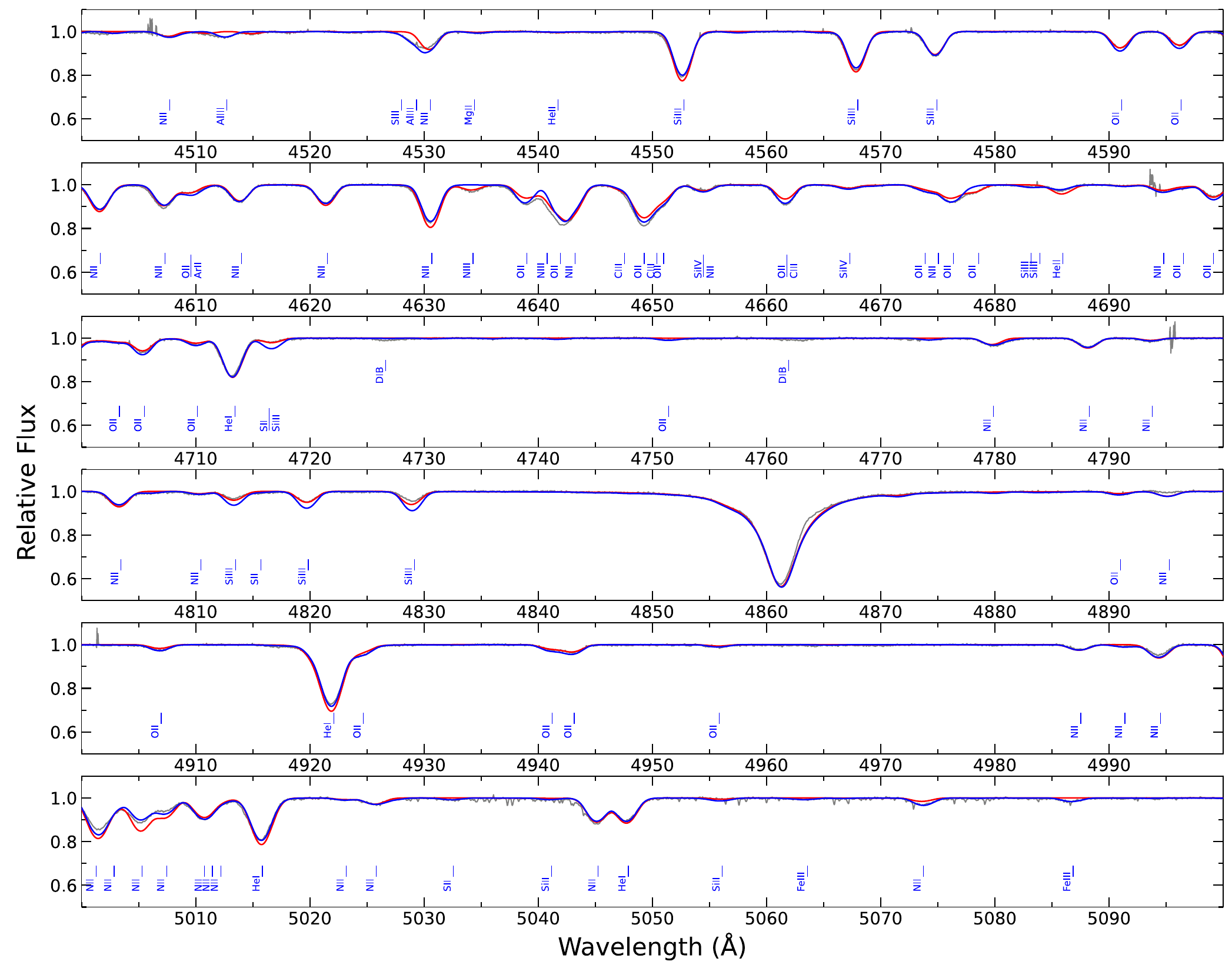}
        \caption{Same as Fig.~\ref{fig:hd93840_1}, but in the wavelength range 4500 to 5100\,{\AA}.}
    \label{fig:hd93840_2}
\end{figure*}

\begin{figure*}
\centering 
\includegraphics[angle=90,width=.98\hsize]{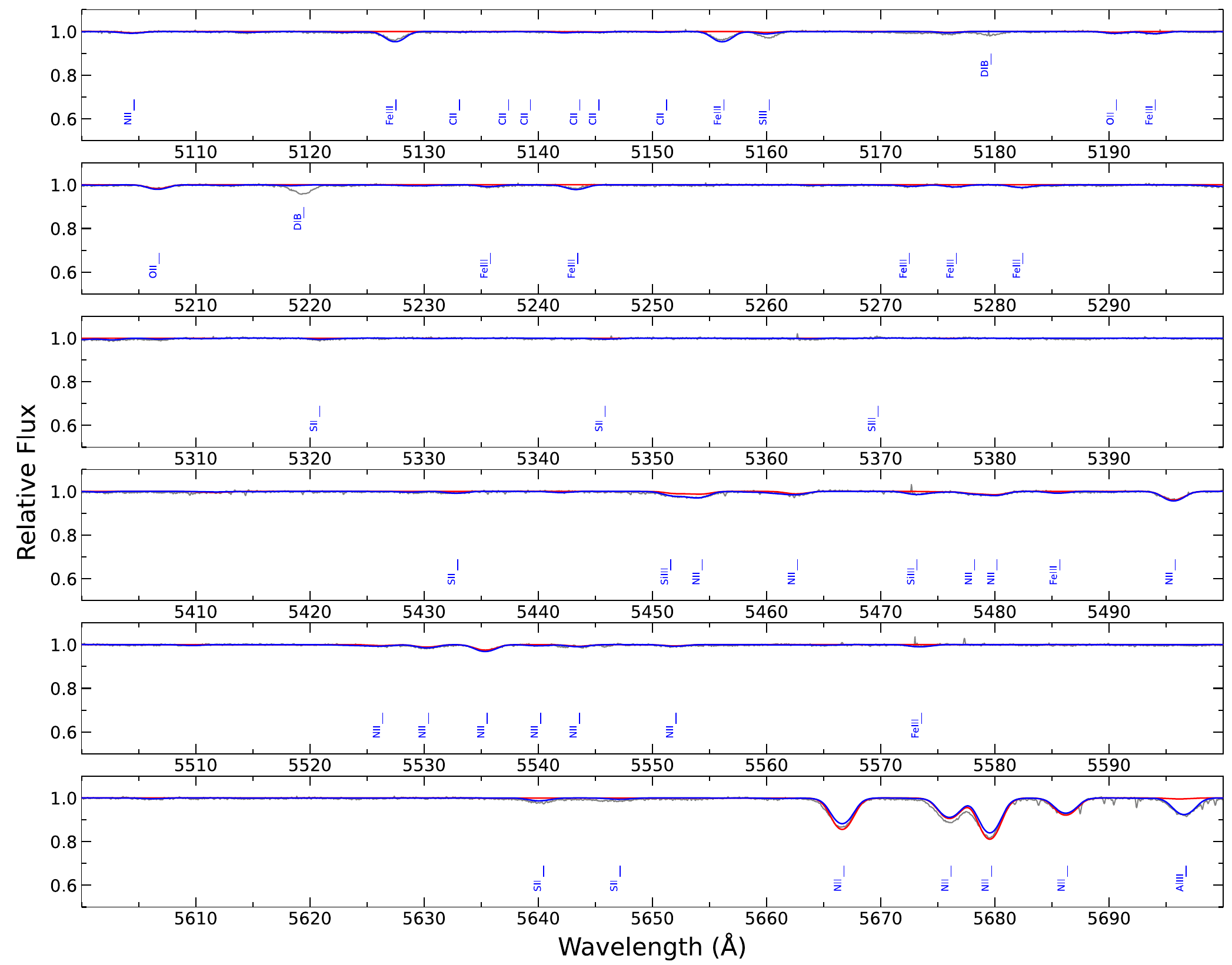}
        \caption{Same as Fig.~\ref{fig:hd93840_1}, but in the wavelength range 5100 to 5700\,{\AA}.}
    \label{fig:hd93840_3}
\end{figure*}

\begin{figure*}
\centering 
\includegraphics[angle=90,width=.98\hsize]{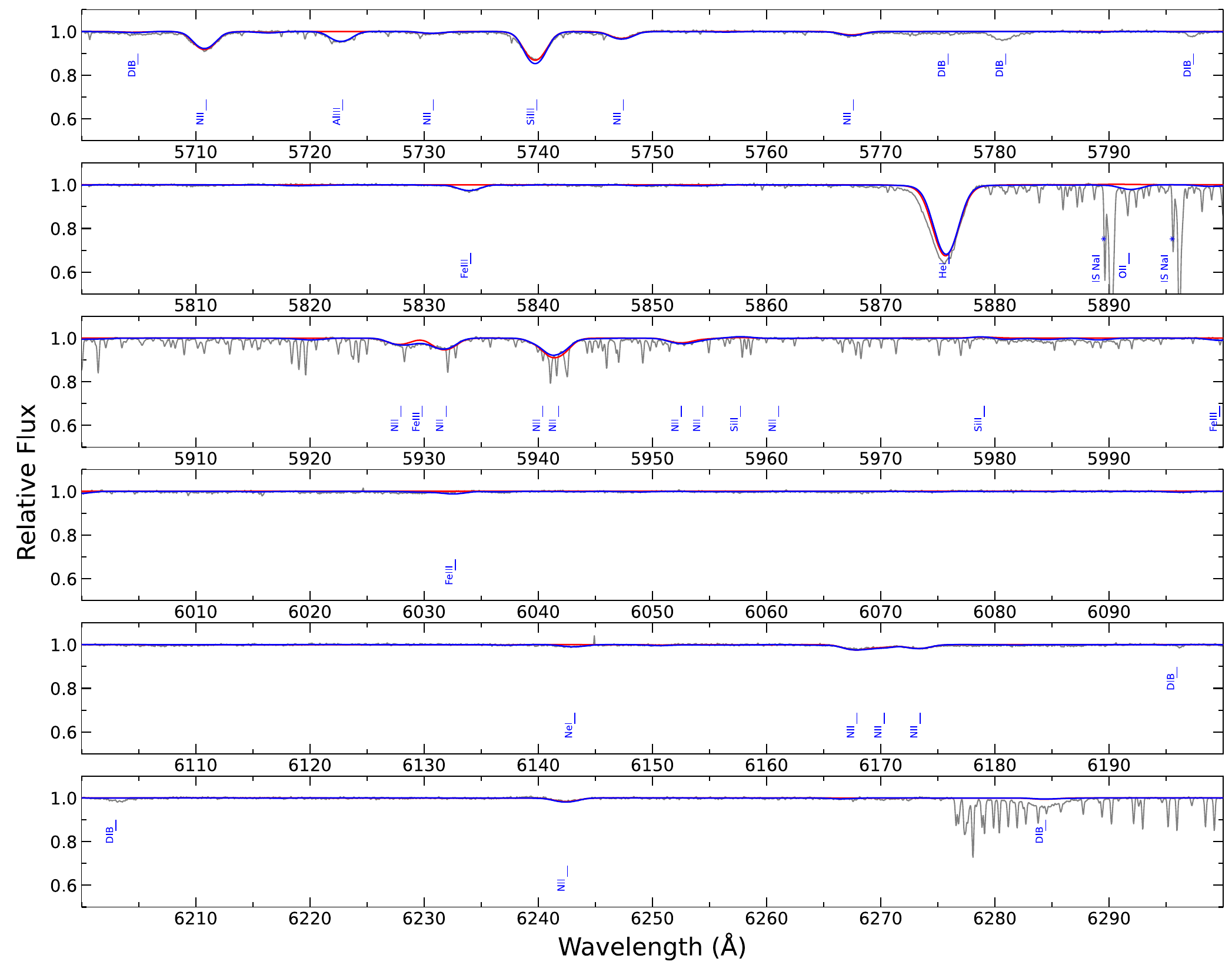}
        \caption{Same as Fig.~\ref{fig:hd93840_1}, but in the wavelength range 5700 to 6300\,{\AA}.}
    \label{fig:hd93840_4}
\end{figure*}

\begin{figure*}
\centering 
\includegraphics[angle=90,width=.98\hsize]{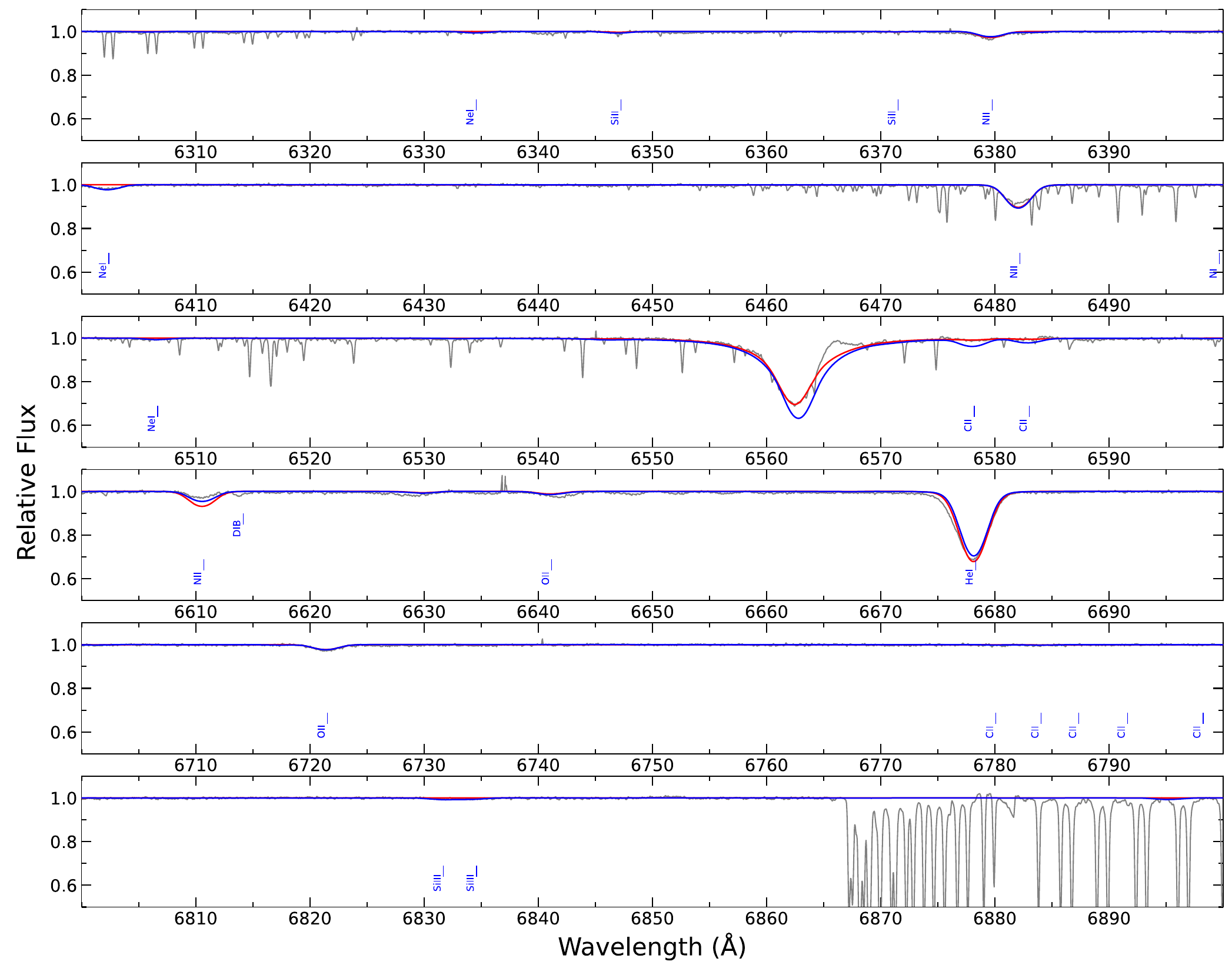}
        \caption{Same as Fig.~\ref{fig:hd93840_1}, but in the wavelength range 6300 to 6900\,{\AA}.}
    \label{fig:hd93840_5}
\end{figure*}

\begin{figure*}
\centering 
\includegraphics[angle=90,width=.98\hsize]{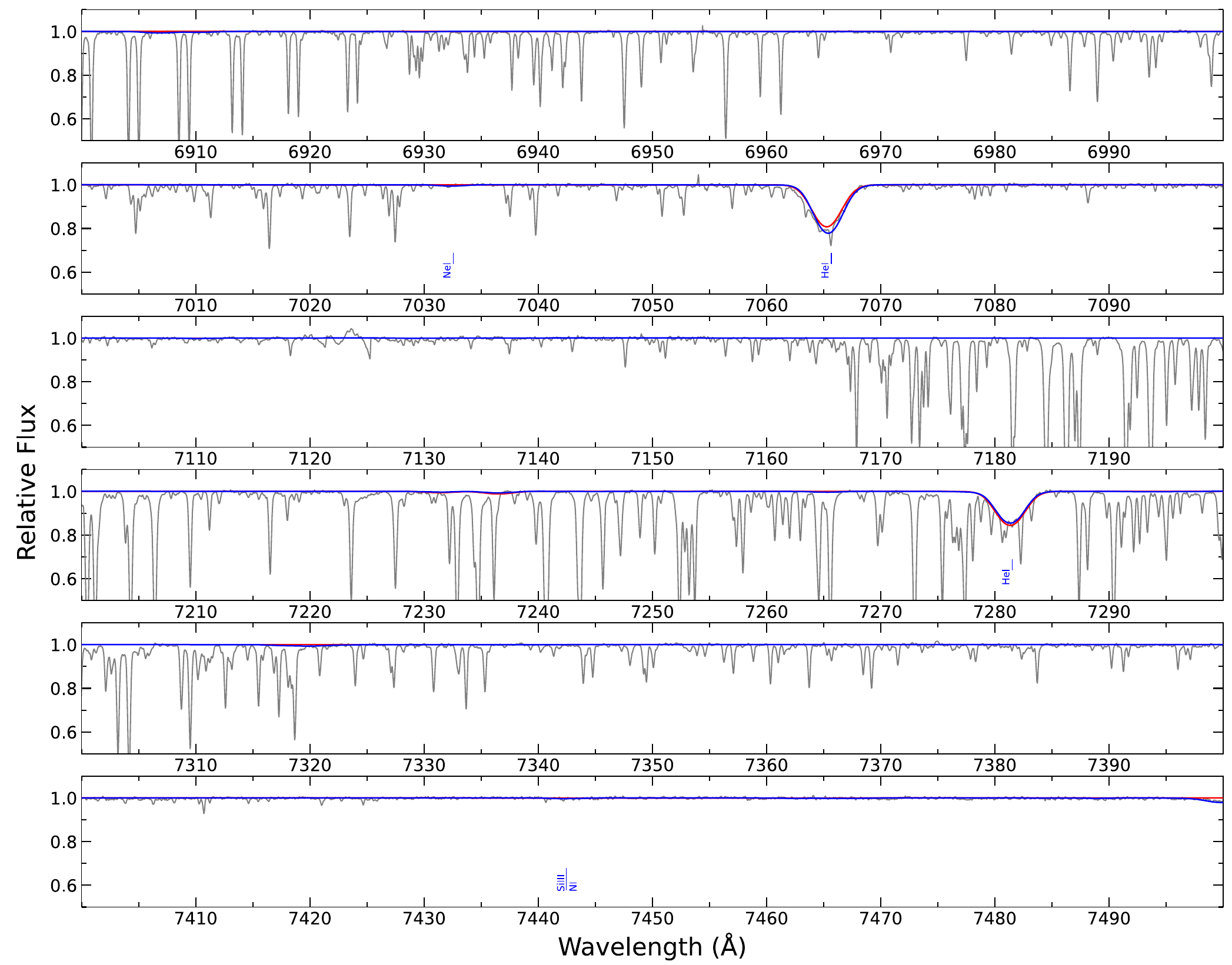}
        \caption{Same as Fig.~\ref{fig:hd93840_1}, but in the wavelength range 6900 to 7500\,{\AA}.}
    \label{fig:hd93840_6}
\end{figure*}

\begin{figure*}
\centering 
\includegraphics[angle=90,width=.98\hsize]{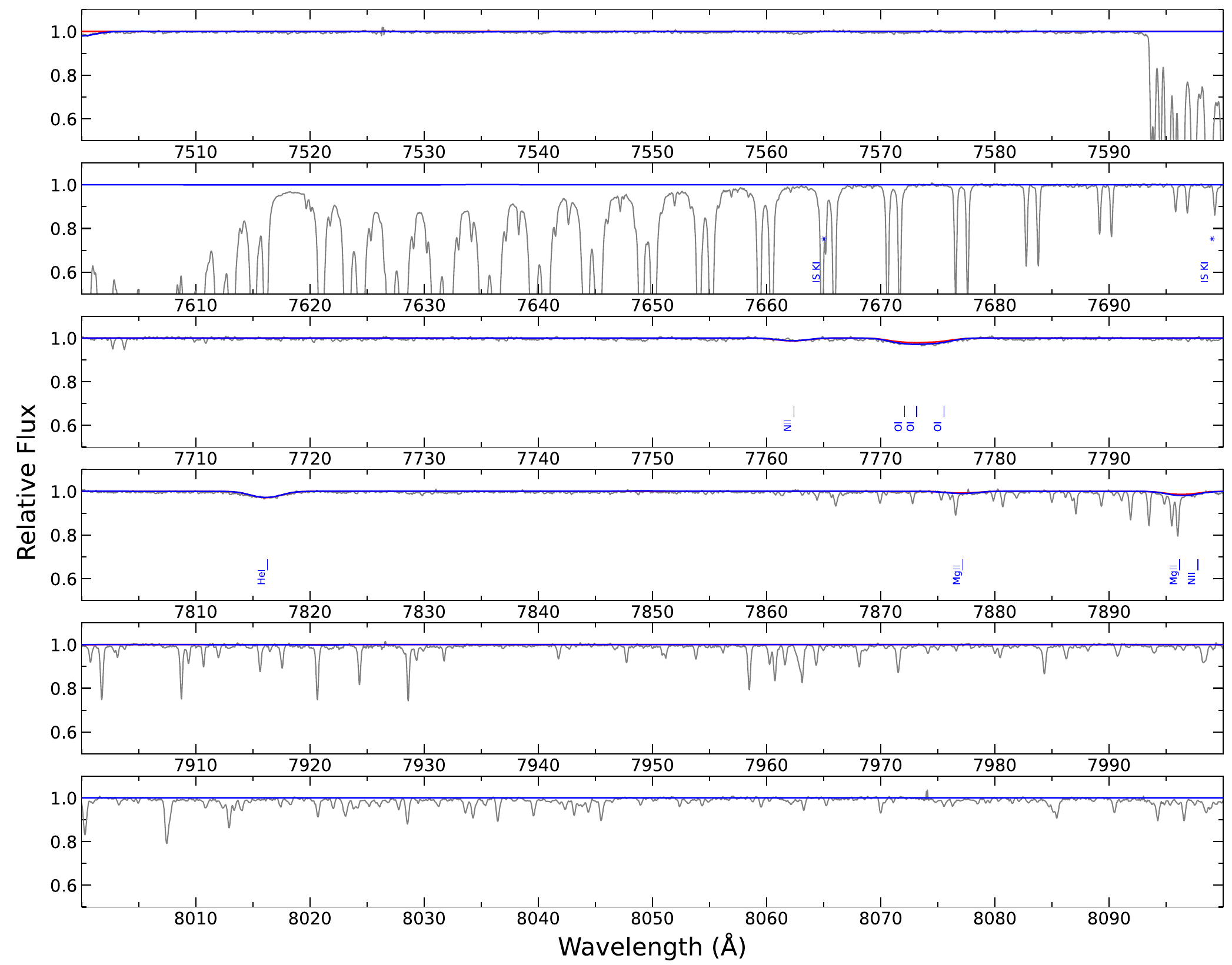}
        \caption{Same as Fig.~\ref{fig:hd93840_1}, but in the wavelength range 7500 to 8100\,{\AA}.}
    \label{fig:hd93840_7}
\end{figure*}

\begin{figure*}
\centering 
\includegraphics[angle=90,width=.98\hsize]{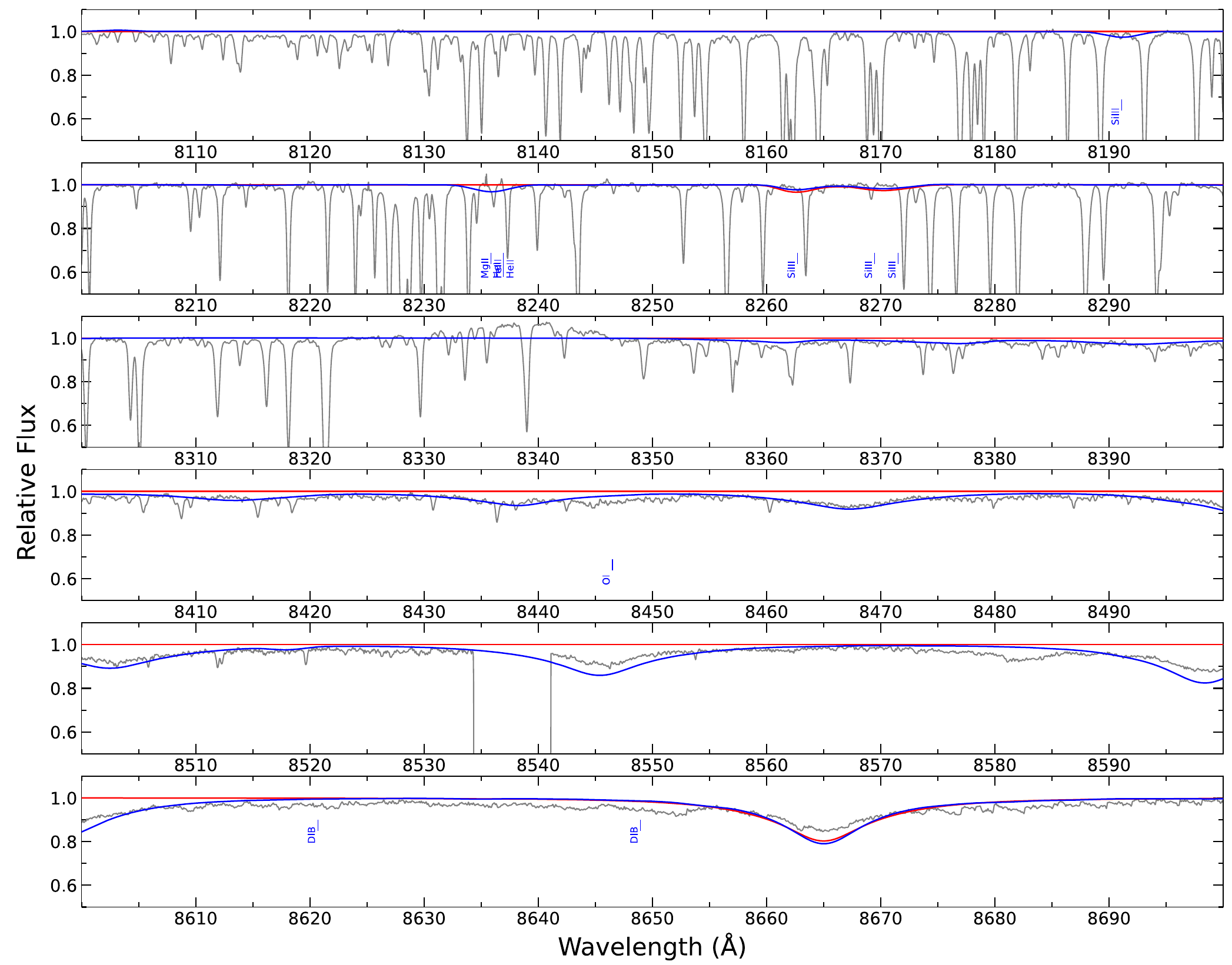}
        \caption{Same as Fig.~\ref{fig:hd93840_1}, but in the wavelength range 8100 to 8700\,{\AA}.}
    \label{fig:hd93840_8}
\end{figure*}

\begin{figure*}
\centering 
\includegraphics[angle=90,width=.98\hsize]{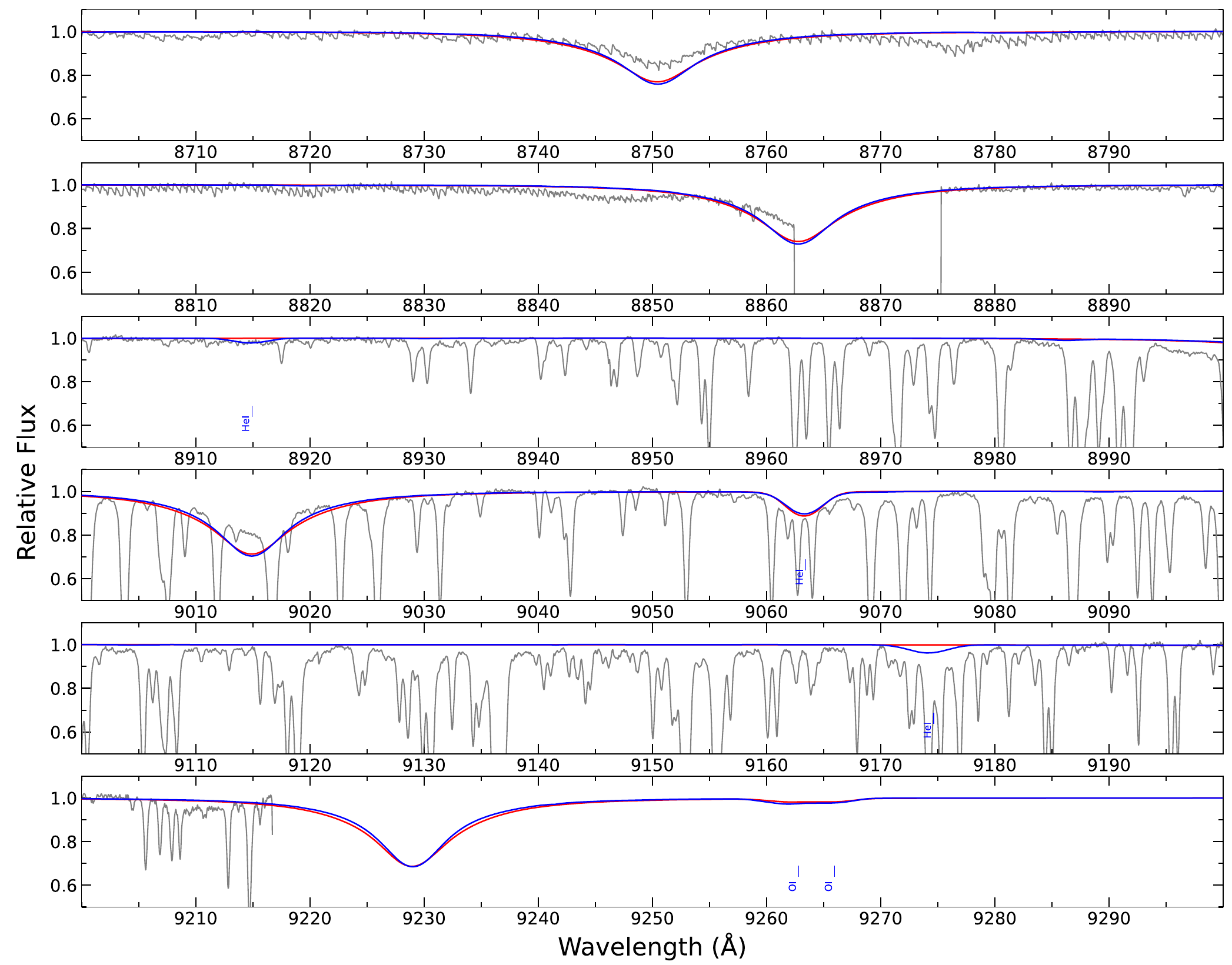}
        \caption{Same as Fig.~\ref{fig:hd93840_1}, but in the wavelength range 8700 to 9300\,{\AA}.}
    \label{fig:hd93840_9}
\end{figure*}

Figures~\ref{fig:hd93840_1} to \ref{fig:hd93840_9} show the comparison of the {\sc Ads} and {\sc Fastwind} model fluxes with the normalised FEROS spectrum. A good match is found overall, with the {\sc Ads} solution reproducing a wider range of spectral lines because more chemical species were considered compared to the {\sc Fastwind} model. We note that \ion{N}{iii} is currently missing in the suite of model atoms used with {\sc Ads}, but it is considered in the {\sc Fastwind} solution. 

 The lower Balmer lines H$\gamma$ and H$\beta$, and in particular H$\alpha$, are slightly affected by the stellar wind, leading to line asymmetries and some filling-in of the line profile by wind emission. Small asymmetries due to extra blue absorption by the stellar wind have also to be noted for the two strongest \ion{He}{i} lines $\lambda\lambda$5875 and 6678\,{\AA}. It turned out not to be straightforward to reproduce the wind signatures of HD~93840. While the wind terminal velocity can rather easily be determined empirically from the presence of well-developed narrow absorption components (NACs) in the strong UV resonance lines to $\varv_\infty$\,$\approx$\,1075\,km\,s$^{-1}$  \citep{Prinjaetal2013}, it
 is not the case for the other wind parameters, in particular the mass-loss rate. The generation of wind emission in high-ionisation species like \ion{N}{V} (requiring $\sim$100\,eV for populating the ion) -- with the \ion{N}{v} $\lambda\lambda$1238.8 and 1242.8\,{\AA} doublet providing the strongest P-Cygni feature in the available UV spectra -- points towards X-ray production in the shocked plasma of a clumped wind. As the focus of the present work is on the photospheric and fundamental stellar properties of the star, we did not pursue this topic further within the {\sc Fastwind} modelling framework.

The wings of higher Balmer lines H$\delta$ and H$\varepsilon$ are matched well, but the comparison of the models with the Paschen series should only be used for guidance, as the normalisation in this spectral region and the quality of the spectrum was complicated by telluric lines and fringing. Concerning ionisation equilibria, in particular the \ion{He}{i/ii}, \ion{O}{i/ii},  and \ion{Si}{ii/iii/iv} (and for the {\sc Fastwind} solution also \ion{N}{ii/iii}) lines are matched simultaneously. Perfect fits for the metal lines could in most cases be achieved by adjusting the abundances within the 1$\sigma$-uncertainties stated in Table~\ref{tab:parameters}, considering that systematic errors of the abundances because of uncertainties in atmospheric parameters, atomic data and continuum placement amount to about 0.1 to 0.15\,dex \citep[e.g.][]{Przybillaetal01b,PrBu01}. Only a few spectral lines show larger deviations, which may be due to model atoms that are rather simple and have seen only minor updates of atomic data for some time, for example \ion{S}{ii/iii} \citep{Vranckenetal96}, or that have a history of being difficult to reproduce such as the \ion{C}{ii} $\lambda\lambda$ 4267, 6578 and 6582\,{\AA} features \citep[][, note that the {\sc Fastwind} implementation of the model atom copes better with these features here]{NiPr06}. The reproduction of the spectrum implies that the detailed atmospheric stratification in the line-forming region is matched well by the model atmospheres.

\begin{figure}[ht!]
\centering
\includegraphics[width=.92\linewidth]{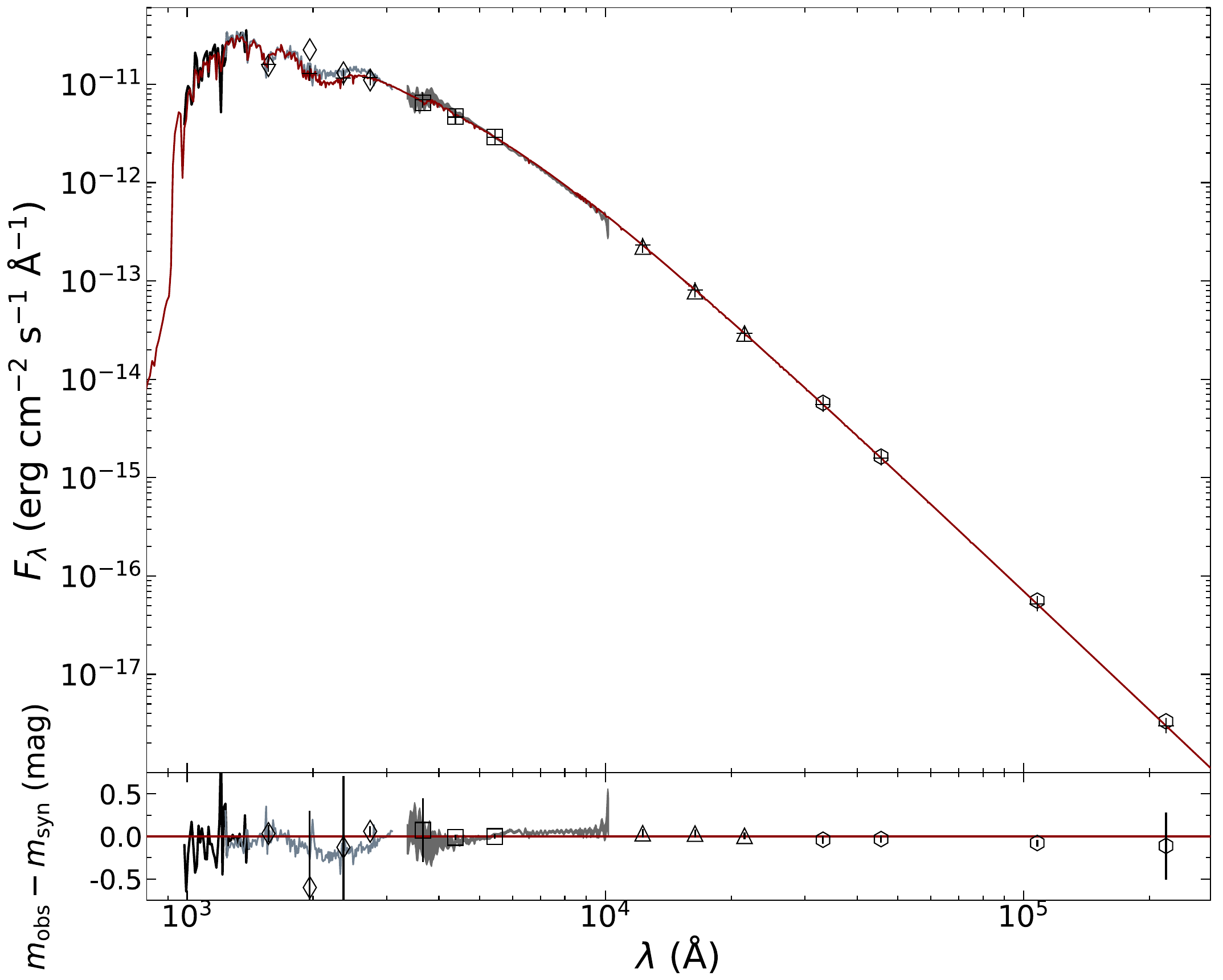}
\caption{Spectral energy distribution of HD~93840. Upper panel: reddened {\sc Atlas} flux (solid red line) compared to various (spectro-) photometric measurements. Lower panel: difference between observed and modelled values in magnitudes. Spectrophotometric data of TUES, IUE, and \textit{Gaia} are depicted as black, light gray, and dark gray lines, respectively. The photometric measurements are illustrated as follows: TD1 (diamonds),  Johnson $UBV$ (squares), 2MASS $JHK$ (triangles) and AllWISE $W1$ to $W4$ photometric data (hexagons). The error bars in the lower panel represent the $2\sigma$ uncertainty ranges. 
\label{fig:sed}}
\end{figure}

Moreover, the models also allow the global energy output of the star from the far-UV below 1000\,{\AA} to the thermal infrared beyond 20\,$\mu$m to be reproduced. Figure~\ref{fig:sed} shows the comparison of the reddened {\sc Atlas} model with the observed SED. A good match with small residuals overall is found, as is also the case for the {\sc Fastwind} model.

Finally, we want to mention the rich spectrum of interstellar lines present in the optical data of HD~93840 (Figs~\ref{fig:hd93840_1} to \ref{fig:hd93840_9}). These cover resonance lines from \ion{Na}{i}, \ion{K}{i} and \ion{Ca}{i/ii}, numerous diffuse interstellar bands (DIBs), and even the molecular cation CH$^+$, which are also identified in the figures.

\section{Distance}\label{appendix:B}
Even in the absence of a direct parallax measurement, one possibility to constrain the distance to a star is the determination of the so-called 'spectroscopic distance'. Based on $T_\mathrm{eff}$ and $\log g$, the stellar 'evolutionary' mass was derived from a comparison with evolutionary tracks for rotating single stars \citep{Ekstroemetal12}, and Eqn.~3 from Paper~I was employed. This provided $d_\mathrm{spec}$\,=\,3740$\pm$340\,pc. The caveat in this is that the evolutionary status of the star has to be reasonably close to that indicated by the stellar evolution models.

Fortunately, the \textit{Gaia} mission \citep{Gaia2016} provided a plethora of parallax measurements $\varpi$. In the literature, the procedure currently most frequently used for deriving distances from the published \textit{Gaia} EDR3 parallaxes is to resort to the data provided by \citet{Bailer-Jones_etal_2021}, derived from a Bayesian approach. 
The 'photogeometric distance' amounts to a (rounded) value of $d_\mathrm{Gaia}$\,=\,2650$^{+240}_{-230}$\,pc (uncertainties are the associated 16$^\mathrm{th}$ and 84$^\mathrm{th}$ percentiles) for HD~93840.
An alternative distance value from the same source may be the 'geometric distance', rounded to 2630$^{+360}_{-240}$\,pc. We note that the distances provided by Bailer-Jones et al. account for the parallax zero-point correction of \citet[, abbreviated as L21 in the following]{Lindegrenetal21}.
These values are in excellent agreement with the inverted \textit{Gaia} EDR3 parallax distance of 2650$^{+270}_{-220}$\,pc, accounting again for the L21 zero-point correction.

\begin{figure}[ht!]
\centering
\includegraphics[width=.92\linewidth]{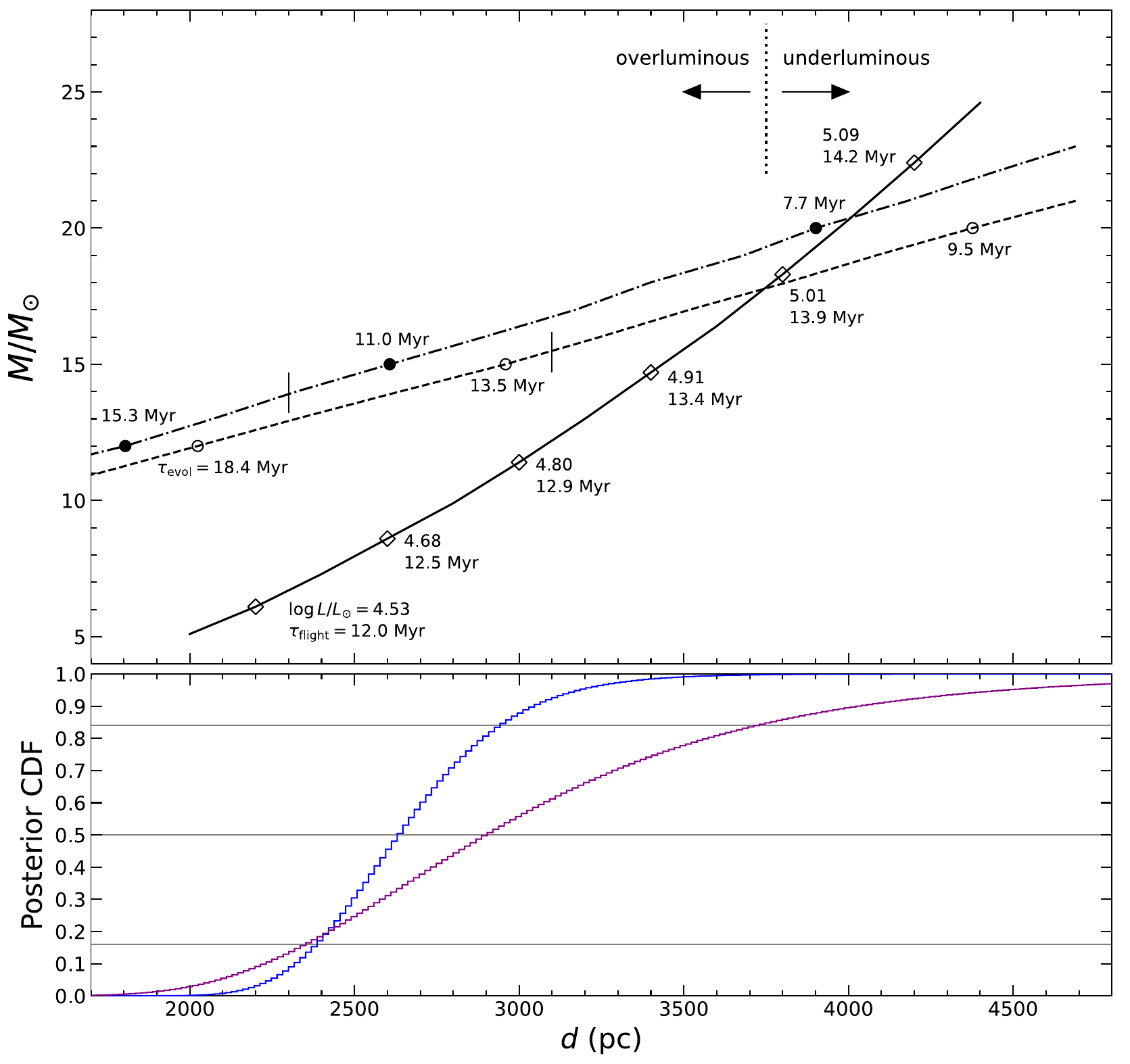}
\caption{Upper panel: the full line follows the loci of stellar masses that reproduce the observed atmospheric parameters ({\sc Ads} solution) of HD~93840 with distance as independent parameter, the dashed and dashed-dotted lines follow the loci where stellar evolution models of \citet{Ekstroemetal12} show the same $T_\mathrm{eff}$ as HD~93840, for the rotating and non-rotating case, respectively. Values along the curves indicate luminosities and flight times $\tau_\mathrm{flight}$ to reach the distance, adopting the kinematic parameters from Table~\ref{tab:parameters} and assuming ejection from the disk mid-plane, or evolutionary timescales $\tau_\mathrm{evol}$ for the models to reach $T_\mathrm{eff}$ of HD~93840. The vertical short lines indicate the points where $\tau_\mathrm{evol}\,=\,\tau_\mathrm{flight}$. The vertical dotted line marks $d_\mathrm{spec}$, as the division line where HD~93840 would switch from being overluminous to becoming underluminous.
Lower panel: posterior cumulative distance distributions (CDF) derived from a Bayesian approach, using the \textit{Gaia} EDR3 parallax after application of the zero-point correction of \citet[][]{MaizApellaniz22}, and assuming an uncertainty inflated according to \citet[][, blue CDF]{ElBadryetal21}, and when considering parallax bias according to \citet[][, purple CDF]{MaizApellaniz22}. The grey horizontal lines represent the 16th, 50th, and 84th percentile.}
\label{fig:distances}
\end{figure}

Yet, the question of the distance to Galactic stars even in the core region of \textit{Gaia} measurements ($\sigma_\varpi/\varpi$\,$<$\,0.1 for HD~93840, making this parallax very precise compared to most of the \textit{Gaia} data) is still under debate with the refined solution of \textit{Gaia} (E)DR3. The issues are the presence and the extent of systematic bias, namely on the parallax zero point and on the parallax uncertainty values.

First, we address the question of the zero-point bias, as this is less critical. For bright blue stars like HD~93840 the parallax zero-point correction -- as determined by \citet{MaizApellaniz22} on basis of blue star cluster members by tying them differentially to the cluster parallax -- may bring improvements over the L21 solution. A more negative zero-point correction is indicated, by a further $\sim$0.012\,mas. This is additionally supported by the independent work of \citet{Flynnetal22} who also found a colour-dependent zero-point offset for bright blue stars of the order $-$0.01\,mas, also based on the analysis of stars in clusters. However, the zero-point difference is small compared to the parallax uncertainty.

\begin{table}[ht!]
\caption{Distance effects on stellar parameters of HD~93840.}
\vspace{-0.6cm}
\label{tab:distanceeffects}
{\footnotesize
\setlength{\tabcolsep}{4mm}
\begin{center}    
\begin{tabular}{ccc}        
\hline\hline
Parameter & {\sc Ads} & {\sc Fastwind}\\ \hline
\multirow{ 3}{*}{$M_V$\,(mag)} & $-5.72^{+0.22}_{-0.21}$ &  $-6.38^{+0.33}_{-0.30}$\\
& $-5.18^{+0.46}_{-0.54}$ & $-5.12^{+0.46}_{-0.54}$\\
& $-4.97^{+0.23}_{-0.25}$ & $-4.91^{+0.23}_{-0.25}$\\[1.5mm]
\multirow{ 3}{*}{$M_{\mathrm{bol}}$\,(mag)} & $-7.75^{+0.23}_{-0.22}$ & $-8.62^{+0.34}_{-0.30}$\\ 
& $-7.20^{+0.46}_{-0.55}$ & $-7.36^{+0.46}_{-0.55}$\\
& $-7.00^{+0.24}_{-0.26}$ & $-7.15^{+0.24}_{-0.26}$\\[1.5mm]
\multirow{ 3}{*}{$M$\,($M_{\odot}$)} & $17.2\pm0.8$ & $22.1\pm1.9$\\ 
& $10.8^{+7.3}_{-3.9}$ & $7.9^{+6.0}_{-3.1}$\\
& $8.9^{+2.8}_{-2.0}$ & $6.5^{+2.8}_{-1.9}$\\[1.5mm]
\multirow{ 3}{*}{$R$\,($R_{\odot}$)} & $22\pm2$    &  $29\pm5$\\ 
& $17^{+5}_{-3}$ & $17^{+5}_{-3}$\\
& $16\pm2$ & $15\pm2$\\[1.5mm]
\multirow{ 3}{*}{$\log L/L_\sun$}    & $5.00\pm0.09$   & $5.34^{+0.12}_{-0.13}$\\ 
& $4.78^{+0.22}_{-0.19}$ & $4.84^{+0.22}_{-0.19}$\\
& $4.70 \pm 0.10$ & $4.76 \pm 0.10$\\
\hline
\end{tabular}
\end{center}}
\vspace{-0.5cm}
\tablefoot{Values are given for the atmospheric parameters derived using {\sc Ads} and {\sc Fastwind}, for three distances: $d_\mathrm{spec}$\,=\,3740$\pm$340\,pc, $d_\mathrm{Gaia}^\mathrm{MA}$\,= $2910^{+830}_{-550}$\,pc and  $d_\mathrm{Gaia}^\mathrm{MA,EB}$\,= $2640^{+310}_{-250}$\,pc (top to bottom).}
\end{table}

Work subsequent to \textit{Gaia} EDR3 found that the published parallax uncertainties may be somewhat underestimated \citep[see e.g. Sect.~3.3 of][]{Gaia2023}. Using the \textit{Gaia} EDR3 parallax, the zero-point correction of \citet{MaizApellaniz22} and the formula of \citet{ElBadryetal21} for the parallax uncertainty adjustment, we derive a distance to HD~93840 of $d_\mathrm{Gaia}^\mathrm{MA,EB}$\,= $2640^{+310}_{-250}$\,pc from a Bayesian analysis using a prior from \citet{MaizApellanizetal2008} to describe the distribution of the OB star population. The corresponding posterior cumulative distance distribution (CDF) is displayed in the lower panel of Fig.~\ref{fig:distances}.
The existence of a much larger parallax bias for relatively bright and blue stars like HD~93840 was proposed more recently by \citet{MaizApellaniz22}. Taking also into account the zero-point  bias of Maiz-Apell\'aniz we derive $d_\mathrm{Gaia}^\mathrm{MA}$\,=\,2910$^{+830}_{-550}$\,pc, again from a Bayesian analysis and stating 16$^\mathrm{th}$ and 84$^\mathrm{th}$ percentiles (see the lower panel of Fig.~\ref{fig:distances}, the flatter CDF). Both studies are solid within the assumptions made, the data adopted and the results obtained, consequently the distance to HD~93840 cannot be tightly constrained on the basis of the \textit{Gaia} EDR3 parallax alone. We therefore consider both solutions in the following. We note that both solutions roughly agree on a minimum distance of about 2.4\,kpc. Moreover, $d_\mathrm{Gaia}^\mathrm{MA,EB}$ is in excellent agreement with the distance values provided by \citet{Bailer-Jones_etal_2021} and the inverted parallax value, despite the (slightly) inflated parallax uncertainty and the enhanced zero-point correction, which mutually have a compensating effect.

In order to attempt to constrain the distance to HD~93840 further, we investigated what consequences the derived distances have for the fundamental stellar parameter determination. Using the atmospheric parameters, extinction, and bolometric correction determined in the model atmosphere analysis, we derived a unique combination of luminosity, radius and mass per given distance value, all of which increase for increasing distance. Table~\ref{tab:distanceeffects} shows some of the fundamental parameter values for our two quantitative spectroscopy solutions ({\sc Ads} and {\sc Fastwind}), considering different characteristic distance values encountered before. The results for the stellar luminosity are visualised in the HRD in Fig.~\ref{fig:hrd}. We want to emphasise that the luminosity of HD~93840 would resemble that of a normally-evolving star only for $d_\mathrm{spec}$. HD~93840 is overluminous for shorter distance values and underluminous for longer distances, with respect to a normally-evolving star of the same mass at the measured~$T_\mathrm{eff}$. 

Further constraints on the distance to HD~93840 can be derived from the masses, and the associated stellar evolution times. This is possible because of the runaway nature of the star, in that the flight time from the star-forming regions of the disk to the current position cannot be longer than the time for the star to evolve to its state in the HRD. While further details of the kinematic calculations in the Galactic potential are discussed in the Appendix~\ref{appendix:C}, we concentrate here already on the conclusions. One finds that the normally-evolving single-star scenario with HD~93840 located at $d_\mathrm{spec}$ can be dismissed. In fact, the range of potential masses can be constrained to below about 14 to 15.5\,$M_\sun$ (depending on initial rotational velocity) for normally-evolving single stars. This is a hard limit assuming dynamical ejection at birth, as H-burning constitutes the longest burning-phase in the stellar life. Overluminous stars in more advanced nuclear burning phases have shorter evolution times. This mass limit and the corresponding evolutionary age imply the distance to HD~93840 to be shorter than a hard limit~of~$\sim$3.1kpc.

However, one can push the constraints further. Assuming for example a current luminosity of $\log L/L_\sun$\,=\,4.68 (for $d$\,=\,2.6\,kpc) to be an upper limit throughout the 12.5\,Myr of time to reach its current position, one can calculate the total energy requirement to an equivalent of fusing about 5\,$M_\sun$ of hydrogen into helium, accounting for a $\sim$10\% additional contribution from 3$\alpha$ burning. As the luminosity likely increased with time, we may assume a mass of 3 to 4\,$M_\sun$ of H turned into He -- plus about less than a Chandrasekhar mass of He-core\footnote{We can again use a lifetime argument here, a more massive He-core would not live long enough for the second mass transfer in the initial binary system to occur.} from the sdO progenitor that has in the meantime been burnt into carbon -- to be more realistic. We note that the luminosity of HD~93840 varies by a factor of less than two in the entire relevant distance range, such that the considerations above should be accurate to within~better~than~1\,$M_\sun$.

Furthermore, we can calculate the overluminosity of HD~93840 as a function of distance, which ranges from a factor over twenty for $d$\,=\,2.4\,kpc to slightly above two at $d$\,=\,3.1\,kpc, implying mean molecular weights (see Sect.~\ref{sec:discussion}) over the entire star of $\mu$\,$\approx$\,1.3 to 0.7, respectively. The $\mu$ at the lower distance value would imply the star to consist almost entirely of helium and heavier elements, which seems unlikely, given the hydrogen-dominated atmosphere. On the other hand, one may calculate the mean molecular weight of the star given the constraints on the carbon core and He-masses above, plus adding mildly processed envelope material accreted from the former companion for the remainder of the total mass derived for a given distance. This provides values of $\mu$ between $\sim$1.05 (for $d$\,=\,2.4\,kpc) and $\sim$0.85 (for $d$\,=\,3.1\,kpc), which means too low values -- compared to the global mean molecular weights from considering the overluminosities -- at the short distance and too high values at the long distance limit. About consistent values are only obtained in the distance range between 2.6 to 2.8\,kpc, which we finally adopt as the most likely distance range to HD~93840. This tight distance distribution is similar to those of the \textit{Gaia}-based distances discussed above, except the one with uncertainties inflated according to \citet{MaizApellaniz22}. Interestingly, for a distance of 2.7\,kpc the overluminosity of HD~93840 is a factor of seven, the average value found by \citet{Senetal22} for the overluminous components in binaries after interaction. 

An independent confirmation of the distance to HD~93840 is nonetheless desirable. This has to await future \textit{Gaia} data releases, when systematics will be better understood and statistical uncertainties will get reduced due to a longer baseline and larger number of measurements.

\section{Kinematics}\label{appendix:C}
We employed the \textit{Gaia} coordinates and proper motion components\footnote{The \textit{Gaia} EDR3 proper motions were corrected for magni\-tude-dependent systematics according to \citet{CGB21}, which turned out to be below the percent level for HD~93840.}, the \textit{Gaia}-based $d_\mathrm{Gaia}^\mathrm{MA,EB}$ as discussed above and alternatively the 'long' $d_\mathrm{spec}$, and the radial velocity as measured from the available spectroscopy as input data for the calculation of the Galactic orbit of HD~93840. The Galactic potential as described by \citet{AlSa91} and the code of \citet{OdBr92} were used for the numerical integration. 

For the radial velocity determination in addition to the two FEROS spectra Phase 3 data of a further spectrum observed on 26 November 2013 with the Ultraviolet and Visual Echelle Spectrograph \citep[UVES,][]{Dekkeretal00} on the ESO Very Large Telescope (VLT) at Paranal in Chile were used. The blue ($R$\,$\approx$\,65\,000) and red-arm ($R$\,$\approx$\,75\,000) UVES spectra were investigated independently\footnote{We note that the UVES data was dropped from the quantitative spectral analysis because of the much lower $S/N$ than obtained in the combined FEROS spectrum.}. Each spectrum was investigated in two ways, {\sc i)} by cross-correlation with the model spectrum, concentrating on spectral regions containing only metal lines, and {\sc ii)} by analysing the Doppler shifts of $>$10 individual unblended metal lines per spectrum. This was done because the cores of the hydrogen lines -- and to a lesser extent the helium lines -- may be affected by the velocity field at the base of the (weak) stellar wind. This yielded a combined $\varv_\mathrm{rad}$\,=\,$-$8.1$\pm$0.5\,km\,s$^{-1}$ (1$\sigma$ standard deviation), as no significant change in radial velocity was found for the FEROS and UVES data, despite them having been taken more than eight years apart. We note that the highest accuracy and precision of the two spectrographs was not achieved, which would have required lamp frames to be taken adjacent to the science observations. A search of the literature on radial velocity measurements for HD~93840 provided only the work of \citet{Feastetal55}, who gave a value of $-$2$\pm$4.5\,km\,s$^{-1}$ from five measurements in the years 1952 and 1953. As super\-giants show wind and also potentially pulsational variability -- which have the potential to yield small radial velocity shifts --, and in view of the instrumental limitations in the 1950s it cannot be decided here whether the small difference between our value of $\varv_\mathrm{rad}$ and that of Feast et al.~point to a SB1 nature of HD~93840, or not. We note, however, that such small differences have no significant effect on the orbit calculations. Nevertheless, radial velocity monitoring of the star is suggested to be undertaken in the future. Given the runaway nature of the star, and the radial velocity stability over the eight year span covered by modern instruments, a breakup of the former binary system appears more probable.

The kinematic study of HD~93840, visualised in Fig.~\ref{fig:kinematics}, yields values of the galactocentric Cartesian velocity components in radial and Galactic rotation direction, and perpendicular to the Galactic plane towards the Galactic north pole, $U$\,=\,$-$83\,km\,s$^{-1}$, $V$\,=\,225\,km\,s$^{-1}$, and $W$\,=\,28\,km\,s$^{-1}$, respectively. The orbital motion brought HD~93840 already more than 500\,pc above the Galactic mid-plane, which corresponds to about ten scale heights of the typical OB star population\footnote{The scale height of the OB star population is small, in the range of about 35 to 70\,pc \citep[see e.g.][, and references therein]{MaizApellaniz01}, compared to scale heights of other stellar populations of the Galactic disk. The scale height for the intermediate-age A-F-type star population ($\lesssim$2\,Gyr) is 150 to 200\,pc \citep[ see e.g][]{Mackerethetal17}, whereas the scale height of the old thin-disk star population is usually discussed in the literature to be about 300\,pc, and more. The supergiant HD~93840 is found far from the usual location of massive stars. From the study of the OB star population distribution one also finds that the Sun is located about 10 to 25\,pc above the Galactic mid-plane, see also e.g. \citet[, and references therein]{MaizApellaniz01}.}. Other stars from our previous work show similar trajectories, as visualised in Fig.~\ref{fig:kinematics}. These are the two ON stars with a binary history, \object{HD 14633} and \object{HD 201345} from \citet{Aschenbrenneretal23}, and \object{HD 91316} ($\rho$~Leo), a likely eccentric binary from Paper~II, probably ejected dynamically. The supergiant HD~93840 shows the most pronounced CNO mixing signature among these, well beyond the values of the two main-sequence ON stars.
\end{appendix}

\end{document}